\documentclass[journal]{IEEEtran}
\usepackage{graphicx} 
\usepackage{amsmath}
\usepackage{amssymb}
\usepackage{url}
\usepackage{hyperref}
\usepackage{subcaption}
\usepackage[switch]{lineno}
\title{Machine learning methods for spectroscopic information recovery under ultrafast photon pileup}
\author{
    Jayson~R.~Vavrek,
    Thomas~D.~MacDonald,
    Yue~Shi~Lai
    \thanks{
        J.R.~Vavrek, T.D.~MacDonald, and Y.S.~Lai are with the Applied Nuclear Physics program at Lawrence Berkeley National Laboratory.

        This work was performed under the auspices of the U.S.\ Department of Energy by Lawrence Berkeley National Laboratory under Contract DE-AC02-05CH11231.
        The project was funded by the U.S.\ Department of Energy, National Nuclear Security Administration, Office of Defense Nuclear Nonproliferation Research and Development (DNN R\&D).
    }%
}
\date{\today}

\begin{document}

\maketitle

\begin{abstract}
    We present methods for recovering spectroscopic information from multiple concurrent photon interactions that would normally be lost due to pulse pileup.
    In particular, we focus on machine learning methods to recover information based on spatial (rather than temporal) energy deposition patterns in position-sensitive detectors.
    We construct two representative problems, namely (1) recovering the fraction of total energy deposition stemming from a monoenergetic signal vs.\ a smooth background; and (2) recovering the signal multiplicity, i.e., the number of interacting photons, in a pure-source-term example.
    In the signal fraction recovery problem, we use 3D convolutional neural networks (CNNs), fully-connected neural networks (FCNNs), a network based on the PointNet++ architecture, and two non-machine-learning methods to estimate the signal fraction in synthetic data when up to~$20$ total piled-up photons are present.
    The CNN, FCNN, and PointNet++ models reconstruct the signal energy deposition fractions with root mean square errors (RMSEs) of $14.5\%$, $18.8\%$, and $16.8\%$ given training datasets that fit in-core, while the classical methods perform poorly and will not improve with additional training data.
    In the multiplicity recovery problem, we demonstrate that, when trained with synthetically-piled-up real Cs-137 data, the 3D CNN architecture can recover the multiplicity with sub-photon RMSE, outperforming non-ML baselines.
    These methods can be adapted to future, more specific photon active interrogation applications, helping to re-enable spectroscopic analyses in those domains.
\end{abstract}

\section{Introduction}
In many photon active interrogation applications, the pulsed time structure of the photon beam can result in peak fluxes that are far too large for spectroscopic detectors to resolve individual photon events.
For instance, ultrafast monoenergetic photon sources (MPS)---such as the BELLA~\cite{ostermayr2020bella, chen2023development} laser plasma accelerator inverse Compton scattering (LPA-ICS) source~\cite{albert2023principles} at Lawrence Berkeley National Laboratory (LBNL)---offer the potential to address several long-standing problems in a variety of nuclear security mission spaces~\cite{geddes2017impact}, but have extremely short pulse widths of ${\sim}10$~fs.
In nuclear treaty verification, nuclear resonance fluorescence (NRF)~\cite{metzger1959resonance, kneissl1996investigation} measurements of spent nuclear fuel or warheads~\cite{hagmann2009transmission, quiter2010nuclear, kemp2016physical, vavrek2018experimental, lan2021rapid, lan2021isotope} offer a much higher signal-to-dose ratio using an MPS compared to traditional bremsstrahlung photon sources, and may be easier to analyze from an information security perspective.
By virtue of their narrow beam widths in both energy and angle, MPS-based systems may also enable one to distinguish mock vs.\ real high explosive (HE) material based on grain structure, and verify the state and type of warhead under verification via microscale radiography.
This sensitivity could similarly be used to determine detailed threat object design information in a nuclear/radiological emergency response scenario, to image dry storage casks in safeguards applications, or to provide high-resolution radiographs~\cite{sun2022review, thornton2024next} of cargo and other dense material configurations for border security applications.

However, while the narrow energy bandwidth is attractive in these mission spaces, the short pulse durations can result in complete pileup within the resolving time of the detector.
Spectroscopic photon detectors traditionally separate the responses from individual incident photons by summing energy depositions within the detector resolving time.
In high-flux applications, particularly those where photons are created by an ultrafast pulsed source, multiple photons can arrive within the resolving time, destroying the ability to perform spectroscopy.
Of course, this problem may also be encountered for example in pulsed linac systems coupled to relatively slow detectors such as high-purity germanium (HPGe).
Similarly, dual-energy x-ray cargo interrogation systems must operate in integration mode due to high rates, leading to a loss of spectroscopy and thus material reconstruction degeneracies~\cite{lalor2024fundamental}.

Instead of separating photons in the time domain, in this work we propose to extract spectroscopic information obscured by ultrafast pileup via energy deposition patterns in the spatial domain.
Higher energy photons tend to deposit energy deeper in the detector and undergo more Compton scattering interactions than lower energy photons, and Compton scattering physics introduces correlations between distant detector regions.
Position-sensitive detectors therefore provide a step towards event separation, but still require the use of reconstruction algorithms for handling multiple concurrent interactions.
In the limiting case where every incident photon undergoes a single-interaction photoelectric absorption, the reconstruction would be trivial, but in reality, multiple incident photons undergoing multiple Compton scattering events produces complicated energy deposition heatmaps where both intra- and inter-event sequencing information is lost.
For small numbers of interactions from a single incident photon, nuclear and medical physics analyses often use maximum likelihood or related methods based on Klein-Nishina physics~\cite{schmid1999gamma, pratx2009bayesian}, but this brute-force approach quickly becomes infeasible as the number of possible interaction sequences rises faster than~$N!$ with multiple incident photons.

Approximate solutions and especially machine learning (ML) methods can provide computationally tractable alternatives, as demonstrated for related problems in, for instance, high energy particle physics~\cite{komiske2017pileup}, medical physics~\cite{nasiri2021deep, lee2023experimental}, and especially nuclear physics~\cite{mayer2021classical, rofors2025low}.
In this work, we test various neural networks for reconstructing spectroscopic information lost to ultrafast pileup via spatial patterns of energy deposition, and compare to several ``classical'' or non-machine-learning methods.
Synthetic data is produced via Geant4 simulation of both bremsstrahlung active background photons and high-energy quasi-monoenergetic source photons impinging on a simplified CdZnTe (CZT) detector model.
These interactions are sampled and combined in order to generate piled-up pulses with varying signal-to-background ratios. 
Using these synthetic piled-up pulses, we evaluate non-ML models (simple depth-threshold and depth-threshold likelihood) and various neural network architectures (fully connected, convolutional, PointNet) to compare performance in terms of the reconstructed signal energy deposition fraction.
We also conduct an experimental demonstration of the multiplicity recovery problem with synthetically-piled-up Cs-137 data from a real CZT detector (the H3D M400i).

This paper is structured as follows---Section~\ref{sec:problem_setting} provides a more detailed problem description, and links to Appendix~\ref{sec:appendix_sequences}, which analyzes the high time complexity of maximum likelihood methods, thereby motivating our ML-based solutions.
Section~\ref{sec:methods} describes the data generation and post-processing steps used to produce piled-up datasets, as well as the various algorithms tested.
Section~\ref{sec:results} presents results and compares performance among the various algorithms, while Section~\ref{sec:discussion} discusses limitations and opportunities for future work.

\section{Problem setting}\label{sec:problem_setting}

\begin{figure*}[!htbp]
    \centering
    \includegraphics[width=1.0\linewidth]{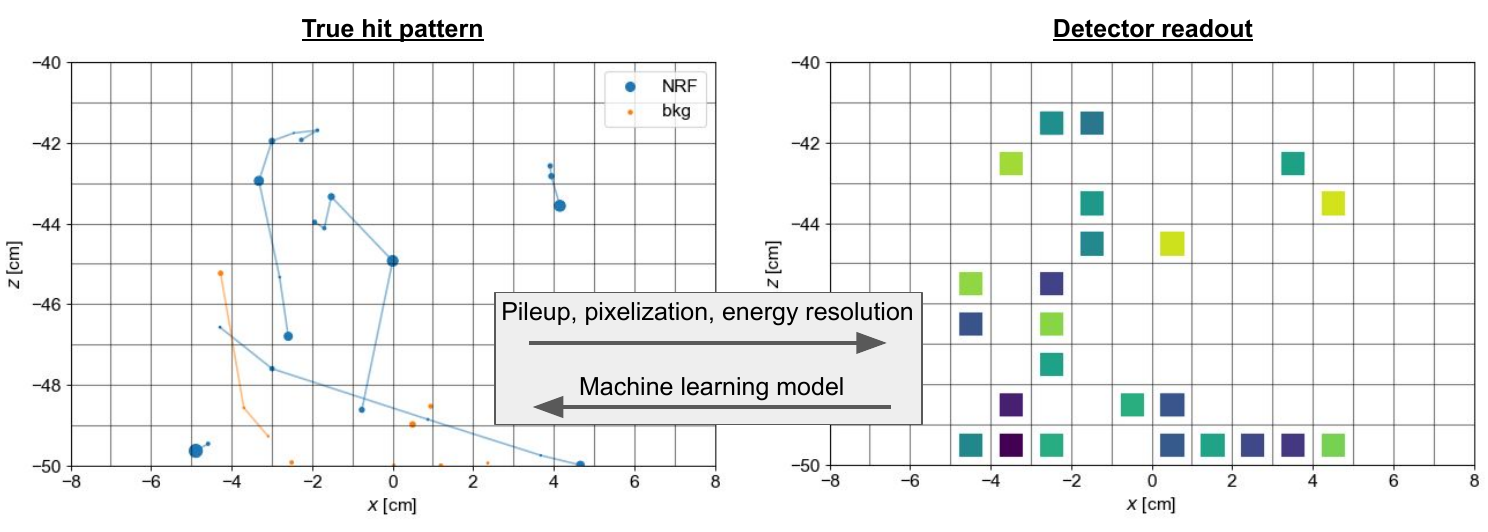}
    \caption{
        Toy model overview of the ultrafast pileup and information recovery problem, using NRF signal photons and bremsstrahlung active background photons incident in the $+z$ direction on a large block of germanium.
        On the left, connecting lines show the true event sequencing and interaction positions are shown as dots; larger dots denote more energy deposition.
        On the right, the interaction positions are discretized and colorized violet to yellow in increasing energy deposition---the \textit{energy deposition heatmap}---and both the intra- and inter-event ordering are lost.
    }
    \label{fig:ultrafast_recon_overview}
\end{figure*}

Here we more formally describe the pileup and spectroscopic information recovery problem.
Assume $K$ photons are incident on a detector---$K_b$ of them from a known background component and $K_s$ of them from a known signal component---and that all their inter-arrival times $\Delta t$ are much smaller than the detector time resolution $\tau$.
Each photon interacts with the detector material primarily through Compton scattering and photoelectric absorption, producing an ordered sequence of interactions or a \textit{true hit pattern}.
$K$ of these true hit patterns are generated essentially simultaneously---the detector cannot resolve the \textit{intra-event} sequencing (ordering of interactions from the same sequence/photon) nor the \textit{inter-event} sequencing (separation of the independent sequences from each photon).
If the detector could perfectly resolve the energy and position of each interaction, it would read out $M$ individual energy depositions $E_m$ and positions $\vec{r}_m$ $\forall\, m = 1, \ldots , M$ (again, with no timing information).
Energy resolution and spatial discretization further degrade the signal to detected energies~$\tilde{E}_n$ and voxelized positions $\vec{q}_n$ $\forall\, n = 1, \ldots , N$, where~$N$, the number of voxels read out, may be smaller than the true number of interactions~$M$ if multiple interactions occur within the same voxel, and is smaller than or equal to the total number of detector voxels~$N_\text{vox}$.
These mechanisms ultimately produce a 3D \textit{energy deposition heatmap} with~$N$ non-zero voxels from which we wish to recover spectroscopic information about the incident flux---see Fig.~\ref{fig:ultrafast_recon_overview}.

Brute-force maximum likelihood techniques for reconstructing full inter- and intra-event sequencing from 3D energy deposition heatmaps rapidly become infeasible as the number of activated voxels~$N$ grows.
Past work~\cite{pratx2009bayesian} has considered~$N$ interactions stemming from $K=1$ incident photons, because other incident photons can most often be time-resolved.
In this case, the number of possible interaction sequences is simply~$N!$.
In the ultrafast pileup case, the number of possible interaction sequences grows faster than~$N!$, since the~$N$ hits may stem from an unknown number of true incident photons~$K$ piled up within the ultrafast pulse, with $1 \leq K \leq N$.
For more detail on the number of possible event sequences, see Appendix~\ref{sec:appendix_sequences}.
As such, completely reconstructing the full inter- and intra-event sequencing via maximum likelihood methods in order to reconstruct spectroscopic information becomes computationally very expensive around $N \gtrsim 10$, which can be reached with even say $K=2$ incident $2$-MeV photons.
However, full sequencing information is not necessarily required---various simpler quantities may be suitable, depending on the application.
In this work, for the purposes of demonstration, we consider two problems, but in future applications, various other quantities may be of interest.

First, we consider a kind of template unmixing problem in which a known monoenergetic photon source and a known broad-spectrum bremsstrahlung background are simultaneously incident on the detector (reminiscent of a transmission NRF measurement), and we wish to recover a single scalar value, namely the fraction~$f$ of energy deposition that stems from the ``signal'' NRF photons.
Here, the information recovery problem is to develop a model $\phi_1$ that produces a signal energy deposition fraction $\hat{f}$:
\begin{align*}
\intertext{{\underline{Problem 1: signal fraction recovery}}}
\textbf{Input:}\quad
\mathbf{x}
   &\in \mathbb{R}^{N_x \times N_y \times N_z}_{\geq 0}\\
\textbf{Output:}\quad
\hat{f} &\in [0,1] \\
\textbf{Model:}\quad
\phi_1 &:\;\,
\mathbf{x}\mapsto \hat f =\phi_1 (\mathbf{x}).
\end{align*}
This problem structure is a natural fit for neural networks, as taken up further in Section~\ref{sec:methods}.
The 3D convolutional neural networks will directly use the shaped data $\mathbf{x} \in \mathbb{R}_{\ge 0}^{N_x \times N_y \times N_z}$, where~$N_x$ is the number of discretizations in the $x$-dimension (and so on), while the fully-connected neural networks will flatten the data shape to a 1D array of length $N_\text{vox} = N_x N_y N_z$.
The objective function to be minimized is the mean squared error (MSE) or equivalently the root mean squared error (RMSE) between the predicted signal fraction~$\hat{f}$ and the true~$f$.
The MSE is a more convenient optimization target as it avoids an extra square-root calculation, while the RMSE is a more physically interpretable quantity, akin to an energy resolution.

Second, we consider the problem of multiplicity recovery, i.e., determining how many incident photons interacted in the detector to produce the observed piled-up signal:
\begin{align*}
\intertext{{\underline{Problem 2: multiplicity recovery}}}
\textbf{Input:}\quad
\mathbf{x}
   &\in \mathbb{R}^{N_x \times N_y \times N_z}_{\geq 0}\\
\textbf{Output:}\quad
\hat{\lambda} &\in \mathbb{R}_{>0}\,. \\
\textbf{Model:}\quad
\phi_2 &:\;\,
\mathbf{x}\mapsto \hat{\lambda} =\phi_2 (\mathbf{x}).
\end{align*}
As written, Problem~2 estimates the continuous Poisson rate parameter $\hat{\lambda}$, which can provide a whole-number photon estimate~$\hat{K}$ via rounding.
It is also written for the total number of interacting photons $K$, but naturally could be extended to predict both $K_s$ and $K_b$ if the two component labels are known in the training data and the network is modified to two outputs.

For historical reasons, the simpler problem of multiplicity recovery is termed Problem~1 rather than Problem~2, and many of the results in this paper focus on the more challenging signal fraction recovery problem.
In addition, due to data availability, experimental demonstrations are restricted to the simpler single-source version of Problem~2, while Problem~1 is addressed with purely synthetic data.
More generally, the parameter space of $\{$problem number, single-/multi-value recovery, model architecture, synthetic/measured data, with/without uncertainty quantification$\}$ is too large to address as a full outer product in this paper, so the results of Section~\ref{sec:results} will provide two concrete examples covering useful combinations.

\section{Methods}\label{sec:methods}

\subsection{Simulations and postprocessing}
In the synthetic signal fraction recovery example, radiation interaction data were simulated using Geant4~\cite{agostinelli2003geant4} with the Livermore low-energy physics lists for both the source and background distributions.
Source photons were simulated as monoenergetic with user-defined energy, typically $2$~MeV, and bremsstrahlung active background photons~\cite{bertozzi2009imaging} were simulated by randomly sampling the photon energy from the distribution
\begin{align}
    p(E) \sim \frac{E_\text{max}}{E} - 1,\quad 0 < E \leq E_\text{max}
\end{align}
with an endpoint energy $E_\text{max} = 2.5$~MeV.
An idealized CZT detector was modeled after the commercially-available H3D M400~\cite{m400_spec_sheet} using a single block of CZT with a $2.171$~cm $\times$ $2.171$~cm face and a depth of $1$~cm.
We note that this operationally-relevant~\cite{dodane2023large} detector model is much thinner than the $10$~cm of germanium used in the Fig.~\ref{fig:ultrafast_recon_overview} schematic, and thus our results are expected to provide a lower bound on performance compared to thicker detectors.
Each simulation directed a parallel beam of $n_\text{photons} = 10^7$ photons uniformly covering the entire square face of the detector, and the true positions and energy depositions of each interaction within the crystal volume were saved to file (along with the true signal energy deposition fraction~$f$) for further postprocessing.

Piled-up pulses were then synthesized by uniformly drawing interaction records for $K_s, K_b \in [0, 10]$ photons each from the source and background datasets.
This uniform sampling is not intended to model any particular distribution of source and background multiplicities, but just to provide coverage over the respective parameter spaces.
Gaussian energy resolution blurring was performed by sampling the measured energy
\begin{align}
    \tilde{E} \sim \mathcal{N}(E, \sigma^2) ,
\end{align}
where the energy resolution $\sigma$ (standard deviation) at energy~$E$ is
\begin{align}
    \sigma(E) = \sqrt{\frac{E}{E_0}} \cdot \sigma_0 , 
\end{align}
where $E_0$ is the reference energy at which the energy resolution of the detector is $\sigma_0$.
For CZT we use $E_0 = 662$~keV and $\sigma_0 \approx 2$~keV~\cite{dodane2023large}.
In this proof-of-concept paper, for simplicity we do not model the asymmetric peak shape of CZT~\cite{doniach1970many, namboodiri1996gamma, li2022peak}, nor charge-sharing among pixels~\cite{kim2011charge, kim2014signal}.
Spatial discretization was then applied, segmenting the CZT crystal volume into $20$ bins in $x$ and $y$ and $50$ bins in $z$, and summing energies within the same voxel.

\subsection{Convolutional neural network}\label{sec:methods_cnn}
The 3-dimensional convolutional neural network was built with TensorFlow Keras.
Data were input as 3-dimensional energy deposition maps with a shape of $(20, 20, 50)$ in $x$, $y$, and $z$ bins.
The hidden layers consisted of three convolutional layers (kernel size $3 \times 3 \times 9$ voxels) each followed by a max pooling layer (pooling size $2 \times 2 \times 2$), the output of the convolutional layers then passed through a flattening layer and a dropout layer ($0.5$ dropout fraction).
The dropout layer was connected to a dense layer with 16 nodes that connected to the output layer with one fully connected node with sigmoid activation.
The structure is summarized in Table~\ref{tab:cnn_structure}.
The model was trained with a learning rate of ${10}^{-4}$ using $120\, 000$ training pulses, $30\, 000$ validation pulses, and a batch size of $32$.
Performance was measured by calculating the RMSE of the predicted signal fraction compared to the true signal fraction for the validation set, using early stopping with a patience of $25$ epochs on the validation loss.

\begin{table}[!htbp]
\centering
\caption{\textsc{3D Convolutional Neural Network Architecture}}
\begin{tabular}{l|c|c|c}
\textbf{Layer Type} & \textbf{Filters/Units} & \textbf{Kernel Size} & \textbf{Activation} \\
\hline
Conv3D & 8 & (3, 3, 9) & ReLU \\
\hline
MaxPooling3D & same & (2, 2, 2) & - \\
\hline
Conv3D & 16 & (3, 3, 9) & ReLU \\
\hline
MaxPooling3D & same & (2, 2, 2) & - \\
\hline
Conv3D & 16 & (3, 3, 9) & ReLU \\
\hline
MaxPooling3D & same & (2, 2, 2) & - \\
\hline
Flatten & - & - & - \\
\hline
Dropout & 0.5 & - & - \\
\hline
Dense & 16 & - & Linear \\
\hline
Dense & 1 & - & Sigmoid \\
\end{tabular}
\label{tab:cnn_structure}
\end{table}

\subsection{Fully connected neural network}
The FCNN was built with TensorFlow Keras and structured with input layers, a variable number of hidden layers, and an output layer.
The input layers, consisting of a flattening and a normalization later, and the output layer with a single node with linear activation, were the same for all FCNNs tested.
The hidden layer architecture was specified at runtime and contained $1$--$3$ layers with $2$--$1024$ nodes per layer.
Training was done with $120\,000$ pulse samples with early stopping monitoring validation loss with a patience of 10 epochs.
The validation set contained $30\,000$ pulses.
Performance was measured by calculating the RMSE of the predicted signal fraction compared to the true signal fraction for the validation set.

\subsection{PointNet++}

As the detector readout is sparsely populated, each pulse of piled-up photons may be more compactly represented as a set of sparse hit coordinates and energy rather than the dense representation of the energy deposition heatmap.
This suggests alternative neural network architectures using graph networks that ingest the energy and coordinates instead of the full detector grid.
PointNet++~\cite{Qi_2017_CVPR,NIPS2017_d8bf84be} is a deep neural network architecture for unordered set of space points.
The invariance to the order of the input is achieved by the max pooling operation, which can be shown to approximate arbitrary unordered functions.
The architecture performs hierarchical clustering to capture features of the set at varying distance scales.
PointNet++ has been well-studied for segmentation and classification of 3D point clouds.
It resembles graph neural networks, with similar strengths and limitations to related graph neural networks~\cite{NEURIPS2021_8ea1e4f9,xu2019powerfulgraphneuralnetworks}.

To train PointNet++, the implementation eqnn-jax~\cite{eqnn-jax} is used on top of the JAX framework~\cite{jax2018github,47008}. Traditionally, PointNet++ is applied on 3-dimensional space points.
As our input contains energy in addition to the spatial hit points, PointNet++ is extended to a 4-dimensional space, where the input is constructed by forming the vector $(x, y, z, E)$. The input is dimensioned to accommodate 100 spatial hits (the observed maximum occupancy in the training and validation set is 70--80), and the unused hit entries are zero-padded.
As we are performing a regression task on the signal fraction~$f$, the classification network output is modified to a vector of $(f, 1-f)$, which resembles a binary (fuzzy) classification.
We largely keep the remaining default graph hyperparameters from eqnn-jax.
The PointNet++ architecture consists of two graph networks with 8 blocks, each block consisting of 6 layers of fully connected network with 256 hidden neurons.
The intermediate results from the graph networks are then fed into the final fully connected block with the hidden layer count $(256, 128, 128, 2, 2)$ leading to the aforementioned output.

For training, a two-pass resampling of the data is used to better equalize the ground truth energy fraction distribution, using 100 bins within $[0, 1]$.
First, the energy fraction distribution is built using a ${\sim}10^5$ pile-up subset.
Then a large dataset is constructed with on-the-fly acceptance-rejection using the inversely observed frequencies of the first pass. 
The same 100 bin histogram is then rebuilt using the this larger sample, and a final acceptance-rejection using this inverse frequencies is performed.

An $L_2$ function over the difference of the $(f, 1-f)$ outputs to the ground truth is applied as the optimization loss function.
The Adam optimizer~\cite{kingma2017adammethodstochasticoptimization} is applied with a learning rate of $3\times 10^{-4}$ over 250 epochs.

\subsection{Classical models}
Classical (non-ML) models were developed to provide a basis against which to evaluate the performance of machine learning models.
Given that brute-force maximum likelihood methods that could leverage spatial correlations from Compton pairs are computationally intractable, the classical approaches only considered the depth profiles of energy distribution.
In particular, it was tested whether a simple depth threshold, $d_0$, dividing energy deposition into front/shallow and back/deep components could be developed and then optimized to make predictions~$\hat{f}$.
Given the front-to-total energy deposition ratio $R_{d_0}$, two extremely simple classical models---linear regression and maximum likelihood---were used to find the correlation between $R_{d_0}$ and the signal fraction $f$.
Due to the simplicity of the models, and the fact that non-ML models largely do not improve with increased data, the non-ML models are not expected to provide good predictions~$\hat{f}$.
Rather, they are merely intended to provide a rough baseline against which the ML models will be compared.

For the regression model, the only parameter that needed to be specified was $d_0$.
Linear regression was performed on $R_{d_0}$ versus $f$ in the training set and the line of best fit used to predict $\hat{f}$ of novel pulses from the measured $R_{d_0}$.
Performance was evaluated by calculating the RMSE between $\hat{f}$ and $f$ on the test set.

The maximum likelihood model is specified by two parameters, $d_0$, and $n_{\text{bins}}$, the number of bins used to generate a two-dimensional histogram of the $R_{d_0}$ versus $f$ that was used to predict $\hat{f}$.
Pulses with zero energy deposition in the detector were masked out as no real measured pulse would have zero energy deposition.
The model was fit by first creating the two-dimensional histogram of $R_{d_0}$ versus $f$ for the training set with $n_\text{bins} = 50$.
Each row of the histogram, corresponding to a given $f$, was normalized by the total number of non-zero energy pulses in $f_{\text{bin}}$, the binned true signal fraction.
This results in each row corresponding to the probability mass function of $R_{d_0}$ for each $f_{\text{bin}}$.
For each $R_{d_0}$ bin, the bin in the column with the highest probability was selected as the signal fraction of maximum likelihood, $\hat{f}$.
To predict $\hat{f}$ of a novel measured pulse, the $R_{d_0}$ of the pulse was computed and binned, and the signal fraction with the maximum likelihood for that bin returned as the prediction.

\subsection{Measured data}\label{sec:methods_meas_data}

Data for the experimental demonstration of multiplicity recovery were acquired using a position-sensitive H3D M400i CZT gamma spectrometer with a $161$-{\textmu}Ci Cs-137 source at a $30$~cm standoff for a dwell time of $13.8$~hours.
Single-photon-event candidates were identified by grouping interactions with identical timestamps, and events with total energy depositions of ${\geq}800$~keV were excluded.
Interaction positions were discretized to $22 \times 22$ pixels and $50$ depth bins.
Training and validation sets were then constructed by synthetically piling up $K \in [1, 10]$ event records uniformly randomly ($120\,000$ and $30\,000$ times, respectively, drawing without replacement within each piled-up sample and from disjoint pools of single-photon events for the training and validation sets).
A 3D CNN architecture similar to that of Section~\ref{sec:methods_cnn} using $3$ layers of $8$ convolutional units each and a softplus final activation was then trained for a maximum of $500$ epochs.
Performance was measured via Poisson loss, and early stopping with a patience of $20$~epochs was applied to the loss in the validation set.

Two new classical methods were established to benchmark the CNN performance in the multiplicity recovery problem.
First, a ``total energy'' method simply divides the total energy deposition by the mean energy deposition from a single $662$~keV photon, about $287$~keV.
Second, a ``voxel count'' method computes the average number of voxels hit per photon from the $K=1$ subset, then divides the observed number of hit voxels by this factor.

\section{Results}\label{sec:results}

\subsection{Signal fraction recovery with synthetic data}

Table~\ref{tab:results_summary} provides a summary of the neural network and non-ML models tested on the signal fraction recovery problem, while Sections~\ref{sec:results_cnn} through \ref{sec:results_classical} provide additional detail on results for each model.

\begin{table*}[!htbp]
    \centering
    \caption{
        \textsc{Summary of models tested for the signal fraction recovery problem}
    }
    \begin{tabular}{c||c|c|c|c}
        Model & Best RMSE & Training time & Architecture & Predicts near $0, 1$ ? \\\hline
        Maximum likelihood & $41\%$ & ${\sim} 1$~s & n/a & $\checkmark, \checkmark$ \\
        Linear regression & $28\%$ & ${\sim} 1$~s & n/a & $\times, \times$ \\
        FCNN & $18.8\%$ & ${\sim} 1$~hr & simple & $\checkmark, \checkmark$ \\
        PointNet++ & $16.8\%$ & ${\sim} 1$~day & complex & $\times, \times$ \\
        3D CNN & $14.5\%$ & ${\sim} 10$~min & moderate & $\times, \checkmark$ \\
    \end{tabular}
    \label{tab:results_summary}
\end{table*}

\subsubsection{Convolutional neural network}\label{sec:results_cnn}

\begin{figure*}[!htbp]
    \centering
    \includegraphics[width=1.0\linewidth]{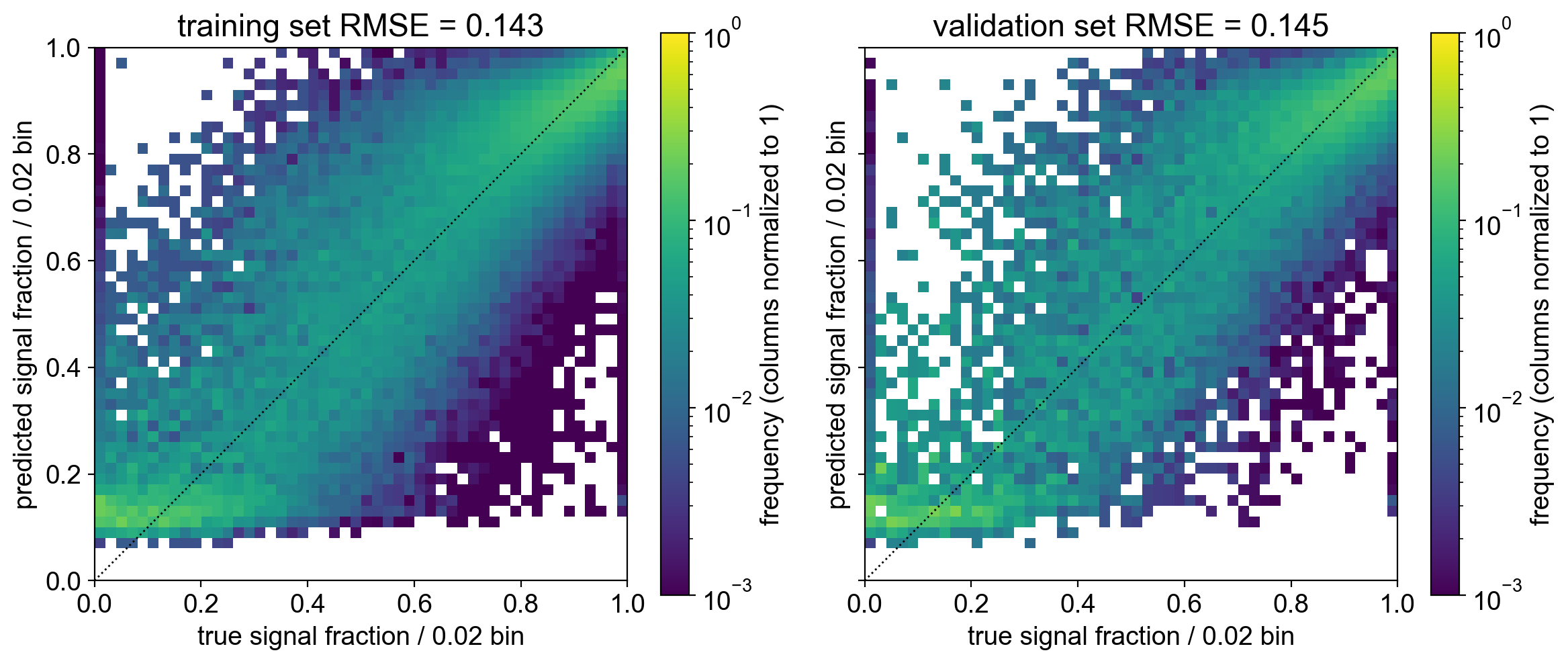}\\
    \includegraphics[width=0.49\linewidth]{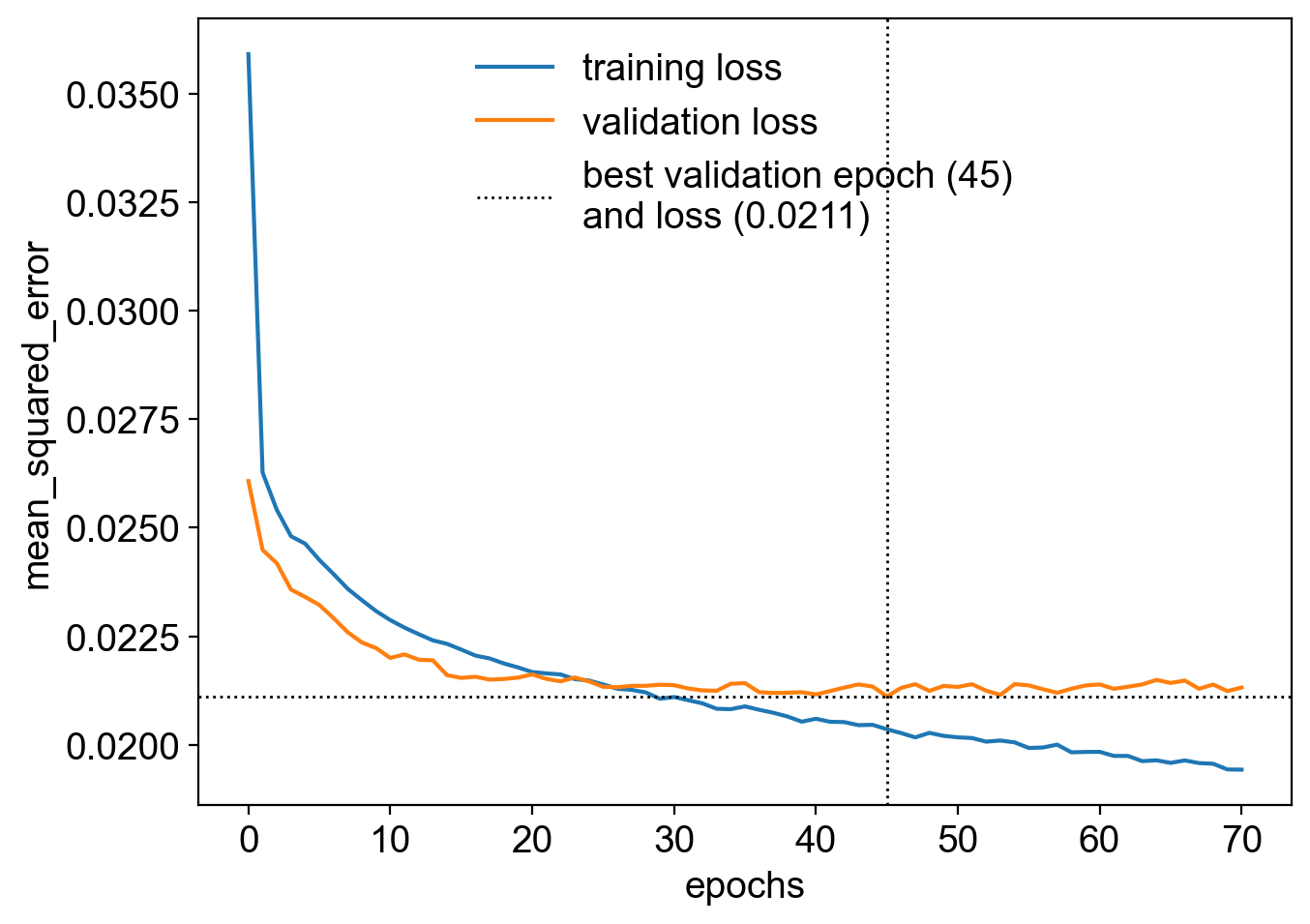}
    \includegraphics[width=0.49\linewidth]{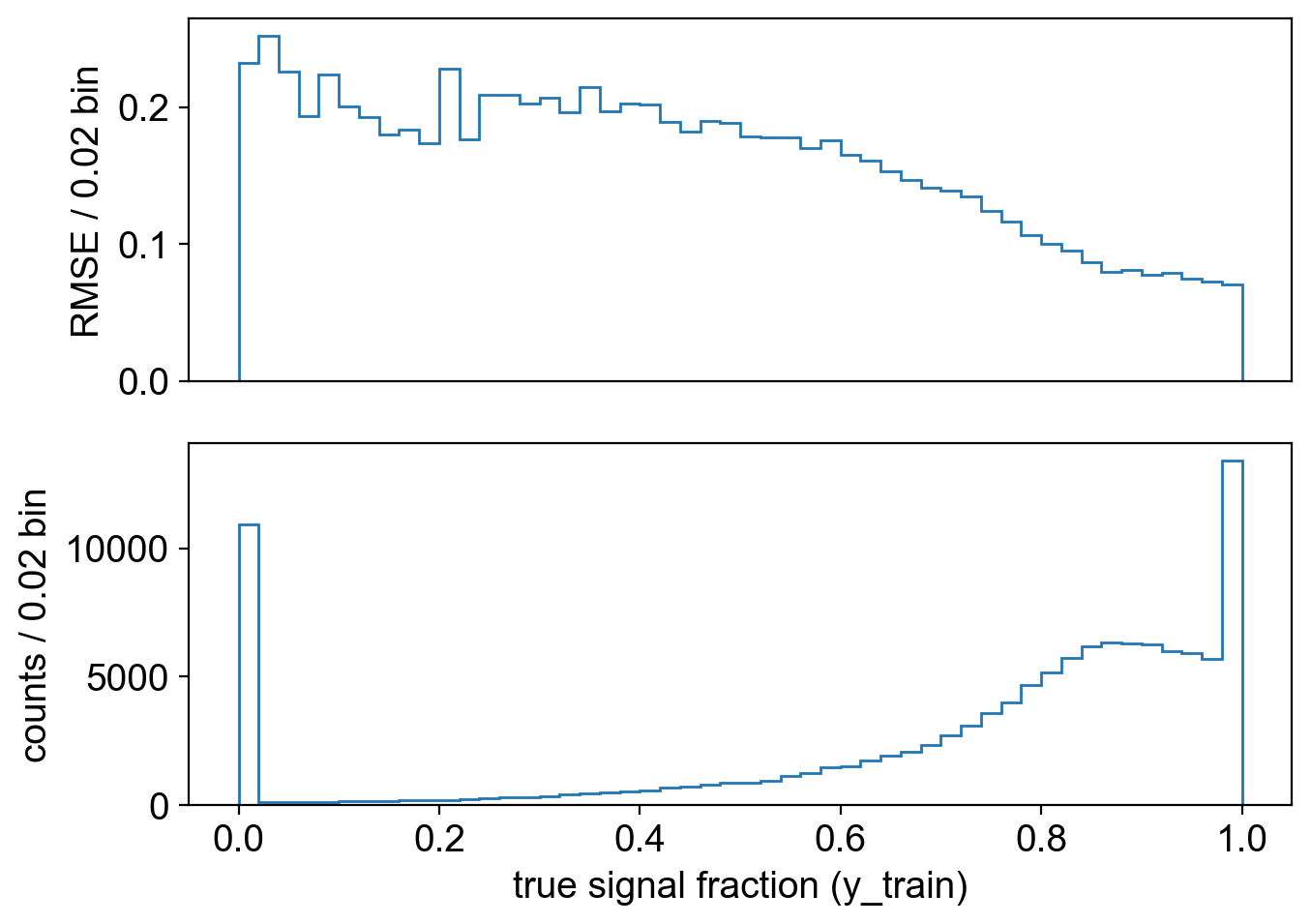}
    \caption{
        Results for the 3D convolutional neural network.
        Top left: training data response matrix.
        Top right: validation data response matrix.
        Bottom left: training and validation losses as a function of epoch.
        Bottom right: RMSE distribution over true signal fractions~$f$, i.e., the $x$-axis projection of the response matrix, and the distribution of true signal fractions~$f$, both for the training set.
    }
    \label{fig:cnn_plots}
\end{figure*}

Fig.~\ref{fig:cnn_plots} shows results from the 3D convolutional neural network.
The CNN achieves a minimum validation loss (MSE) of~$2.11 \times 10^{-2}$, i.e., an RMSE of $14.5\%$, at epoch $45$ ($0$-indexed).
The response matrices generally follow a modest distribution around the $1{:}1$ diagonal line, with one notable exception---similar to the $1024$-node single layer FCNN, the CNN fails to predict any signal fractions lower than~$\hat{f} \approx 0.1$.
This may be a limitation of the training data, which, due to the uniform sampling of $K_s, K_b \in [0, 10]$ signal and background photons, has very few examples of pulses with~$f \lesssim 0.1$.
The distribution of RMSE over true signal fractions~$f$ shows that recovery performance tends to improve in regions with more training data.
Uncertainty quantification (UQ) is addressed in Appendix~\ref{sec:appendix_uq}.
Training the CNN with $120\,000$ training pulses took ${\sim}7$~minutes on an NVIDIA GeForce RTX 4090 GPU, while inference on the training and validation sets took ${<}10$~s combined.

\subsubsection{Fully connected neural network}

A range of FCNN architectures were explored following the design principle of utilizing the smallest model possible to fit the data to avoid overfitting.
Table~\ref{tab:fcnn_results} shows the RMSE of each model when predicting the signal fraction of the validation set data.

\begin{table}[h]
\centering
\caption{
    \textsc{Fully connected neural network architectures and performance results}
}
\begin{tabular}{c|c|c}
layers & nodes & RMSE [\%] \\
\hline
1 & 2 & 24.0 \\
1 & 8 & 23.0 \\
1 & 128 & 23.7 \\
1 & 512 & 21.7 \\
1 & 1024 & 21.5 \\
2 & 4, 2 & 23.0 \\
2 & 16, 4 & 21.6 \\
2 & 128, 16 & 19.3 \\
2 & 256, 64 & 18.8 \\
3 & 8, 4, 2 & 23.8 \\
\end{tabular}
\label{tab:fcnn_results}
\end{table}

\begin{figure*}[!htbp]
    \centering
    \begin{subfigure}[b]{0.31\textwidth}
        \centering
        \includegraphics[width=\textwidth]{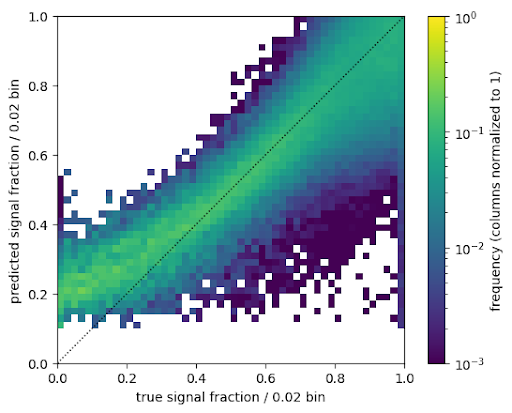}
        \caption{}
        \label{fig:subfig_a}
    \end{subfigure}
    \hfill
    \begin{subfigure}[b]{0.31\textwidth}
        \centering
        \includegraphics[width=\textwidth]{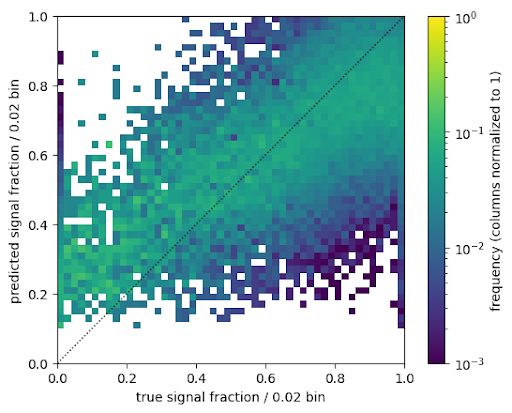}
        \caption{}
        \label{fig:subfig_b}
    \end{subfigure}
    \hfill
    \begin{subfigure}[b]{0.31\textwidth}
        \centering
        \includegraphics[width=\textwidth]{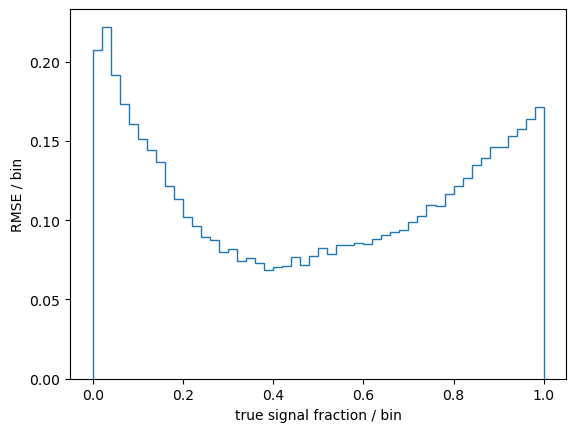}
        \caption{}
        \label{fig:subfig_c}
    \end{subfigure}
    
    \vspace{0.5cm}
    
    \begin{subfigure}[b]{0.31\textwidth}
        \centering
        \includegraphics[width=\textwidth]{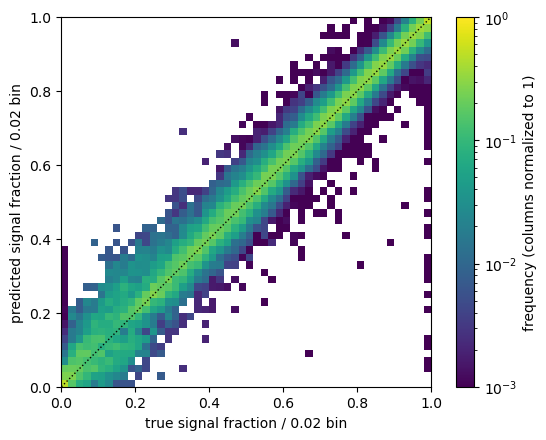}
        \caption{}
        \label{fig:subfig_d}
    \end{subfigure}
    \hfill
    \begin{subfigure}[b]{0.31\textwidth}
        \centering
        \includegraphics[width=\textwidth]{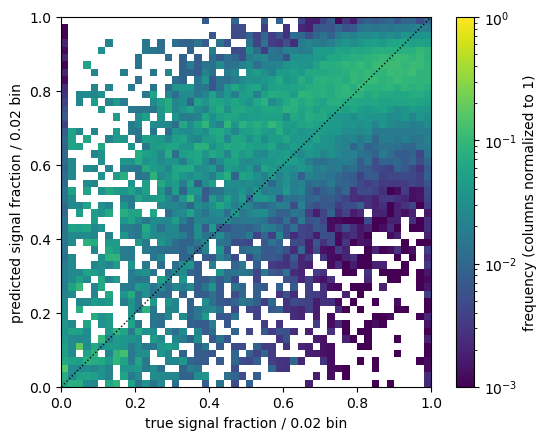}
        \caption{}
        \label{fig:subfig_e}
    \end{subfigure}
    \hfill
    \begin{subfigure}[b]{0.31\textwidth}
        \centering
        \includegraphics[width=\textwidth]{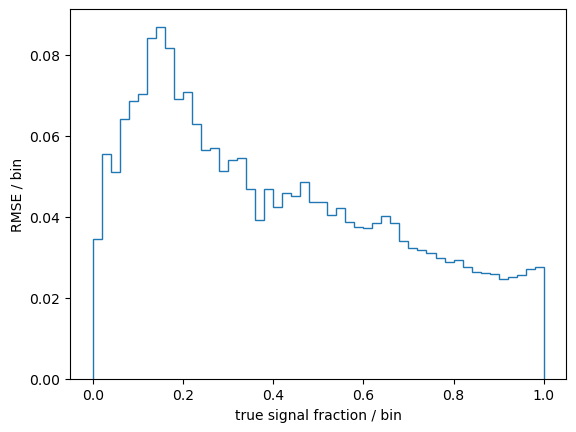}
        \caption{}
        \label{fig:subfig_f}
    \end{subfigure}
    
    \caption{
        Results for the best performing single- and two-layer fully connected neural networks.
        (Top row) $1024$-node single-layer FCNN trained with $n_\text{samples} = 120\,000$.
        (a) True signal fraction~$f$ versus predicted signal fraction~$\hat{f}$ for the training set.
        (b) True signal fraction~$f$ versus predicted signal fraction~$\hat{f}$ for the validation set.
        (c) RMSE for each true signal fraction bin.
        (Bottom row) the same plots for the two-layer architecture with $256$ nodes in the first layer and $64$ nodes in the second.
        In plots (a), (b), (d), and (e), the diagonal dotted line indicates a perfect reconstruction.
    }
    \label{fig:fcnn_plots}
\end{figure*} 

Fig.~\ref{fig:fcnn_plots} shows the results for the best performing single- and two-layer FCNNs.
Correlations between the predicted signal fraction~$\hat{f}$ and true signal fraction $f$ are shown as response matrices.
The 1024 node single layer model achieved an RMSE of $21.5\%$ after training for 3 epochs.
The single layer FCNN response matrix shows that predictions are distributed about the $1{:}1$ diagonal but deviate at low signal fractions such that the model does not predict signal fractions below approximately 0.15.
The single-layer model performs best in the middle range of true signal fractions, where there are a mixture of signal and background photons, and performs worse at the extremes of the signal fractions where pulses are dominated by photons from a single source, as showing in Fig.~\ref{fig:subfig_c}.

The two-layer model achieved an RMSE of $18.8\%$ after training for 18 epochs.
It shows a much tighter distribution about the $1{:}1$ diagonal than the single-layer model on the training data.
It also is capable of predicting the full range of signal fractions unlike the single-layer model.
The response matrix for the validation set (Fig.~\ref{fig:subfig_e}) shows a larger spread than the training set, indicating that the model may be overfitting the training data.
Fig.~\ref{fig:subfig_f}, showing the RMSE on the validation set for discretized $f$, shows that the model performs best at high signal fractions and when the pulse is from pure source or pure background photons.
The RMSE per bin is highest for mixed-source pulses where the signal fraction is low, which is the least well-represented type of pulse in the training set, so this may be improved by utilizing more training data.

\subsubsection{PointNet++}

\begin{figure*}[!htbp]
    \centerline{\includegraphics[width=0.5\textwidth]{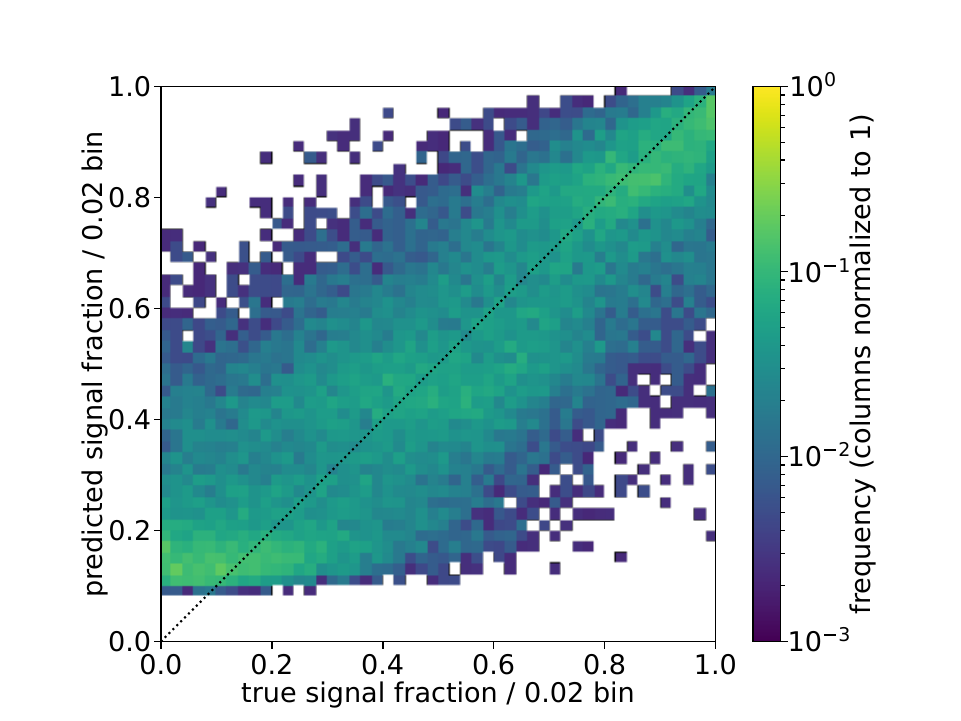}
    \includegraphics[width=0.5\linewidth]{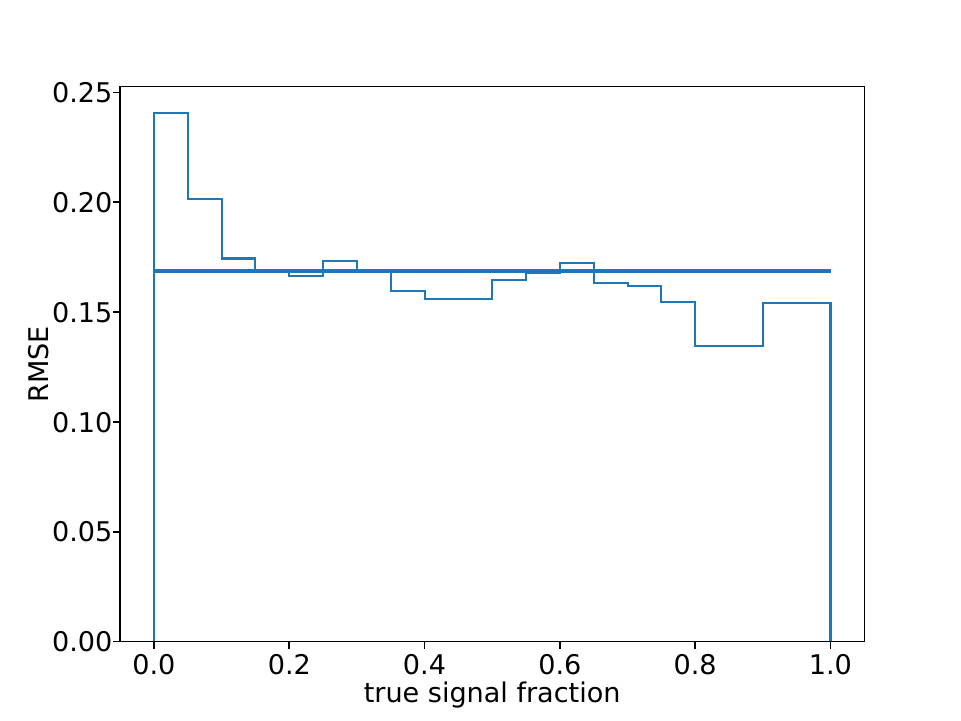}}
    \centerline{\hfill (a)\hfill\hfill (b)\hfill}
    \centerline{\includegraphics[width=0.5\textwidth]{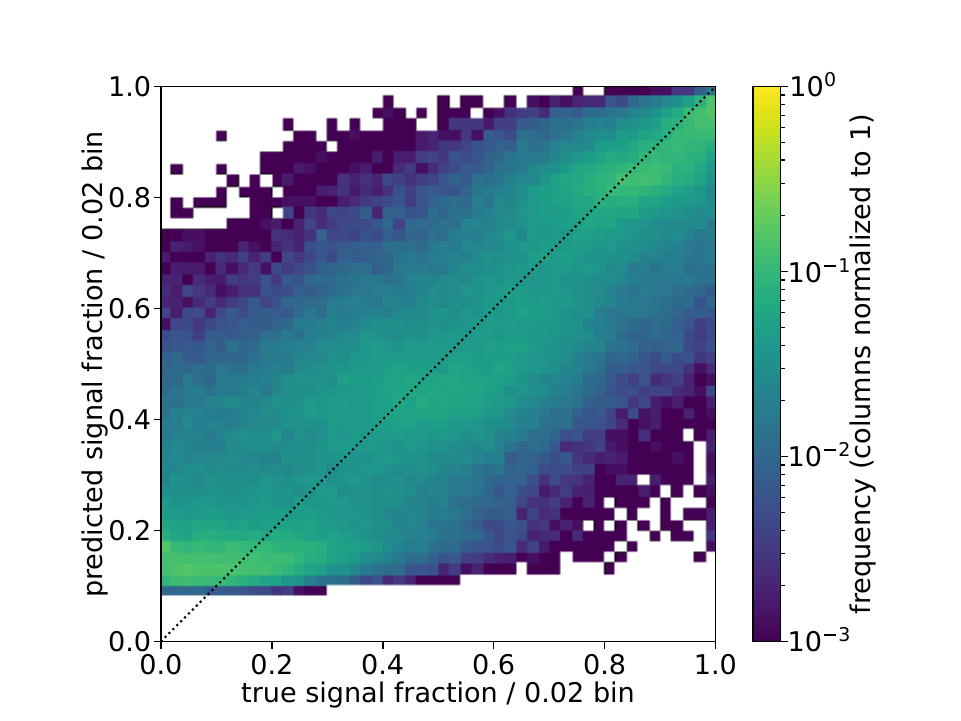}
    \includegraphics[width=0.5\linewidth]{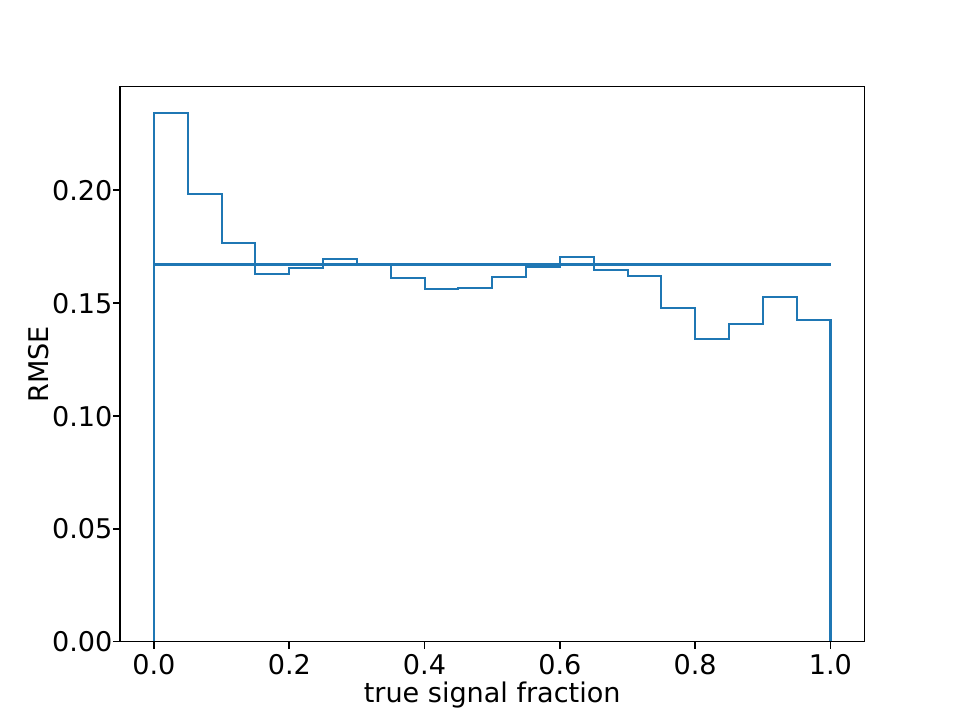}}
    \centerline{\hfill (c)\hfill\hfill (d)\hfill}
    \caption{
        Results for the PointNet++ model.
        (a): training response matrix, (b): training RMSE distribution over true signal fractions~$f$, i.e., the per-bin residual of the response matrix response, (c): validation response matrix, (d): validation RMSE distribution.
    }
    \label{fig:pointnet_plots}
\end{figure*}

Fig.~\ref{fig:pointnet_plots} shows the results from PointNet++ after 50 epochs.
PointNet++ achieves an RMSE of $\approx 0.168$.
The overall RMSE is distributed relatively evenly across the true energy fraction~$f$, unlike the CNN and FCNN, due to the additional event sampling scheme.
Similar to the CNN and FCNN, the response matrix follows the diagonal line between signal fractions of $\approx 0.2$ to $0.9$, and plateaus beyond these values.
The response also exhibits a behavior split roughly into three signal fraction regions.
This in turn corresponds to the residual oscillation visible in the RMSE structure.
For the PointNet++ the saturation of the reconstructed energy fraction to $\approx 0.1$ and $0.9$ can be explained by the overall observed resolution of $\approx 0.2$ and optimizer possibly having to sacrifice the RMSE at $\approx 0.2$ and 0.8, if one would purely optimize for a better RMSE at the energy fraction endpoints.

\begin{figure}[!htbp]
    \centerline{\includegraphics[width=0.5\textwidth]{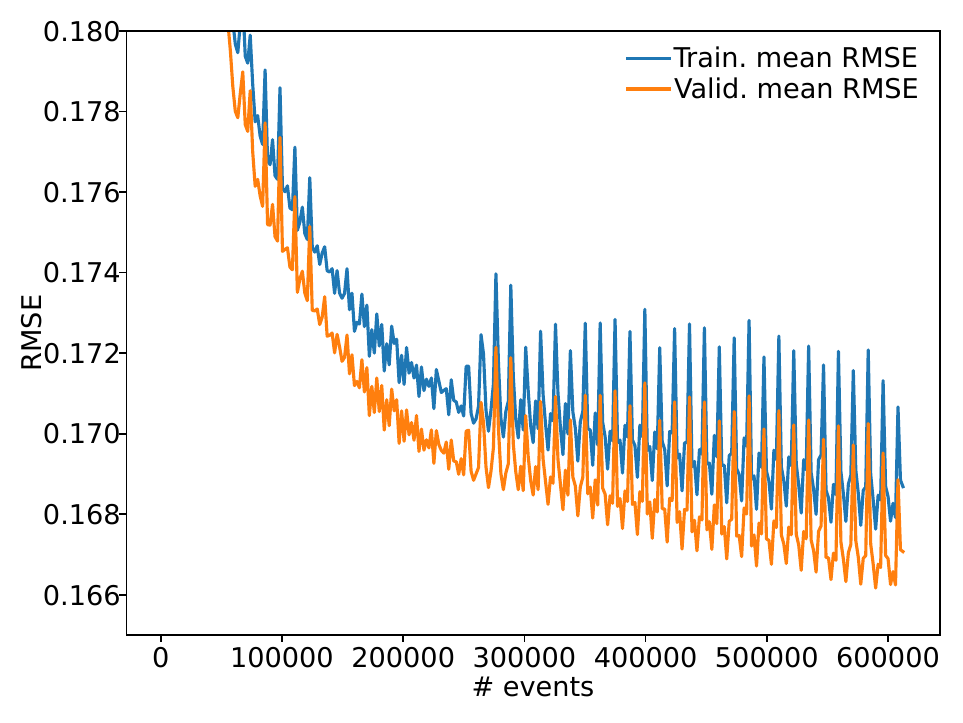}}
    \caption{
        Training and validation RMSE as the function of the training (in number of events) for PointNet++. Note that the change in the RMSE within a training epoch is plotted, and the oscillation visibly matches the batch-to-epoch ratio. It is expected that within an epoch there will be periods of statistical overfitting (relative to the full validation set, which the RMSE evaluation is based on) before the full epoch is seen.
    }
    \label{fig:pointnet_training}
\end{figure}

\subsubsection{Classical models}\label{sec:results_classical}
\begin{figure*}[!htbp]
    \centering
    \includegraphics[width=0.335\linewidth]{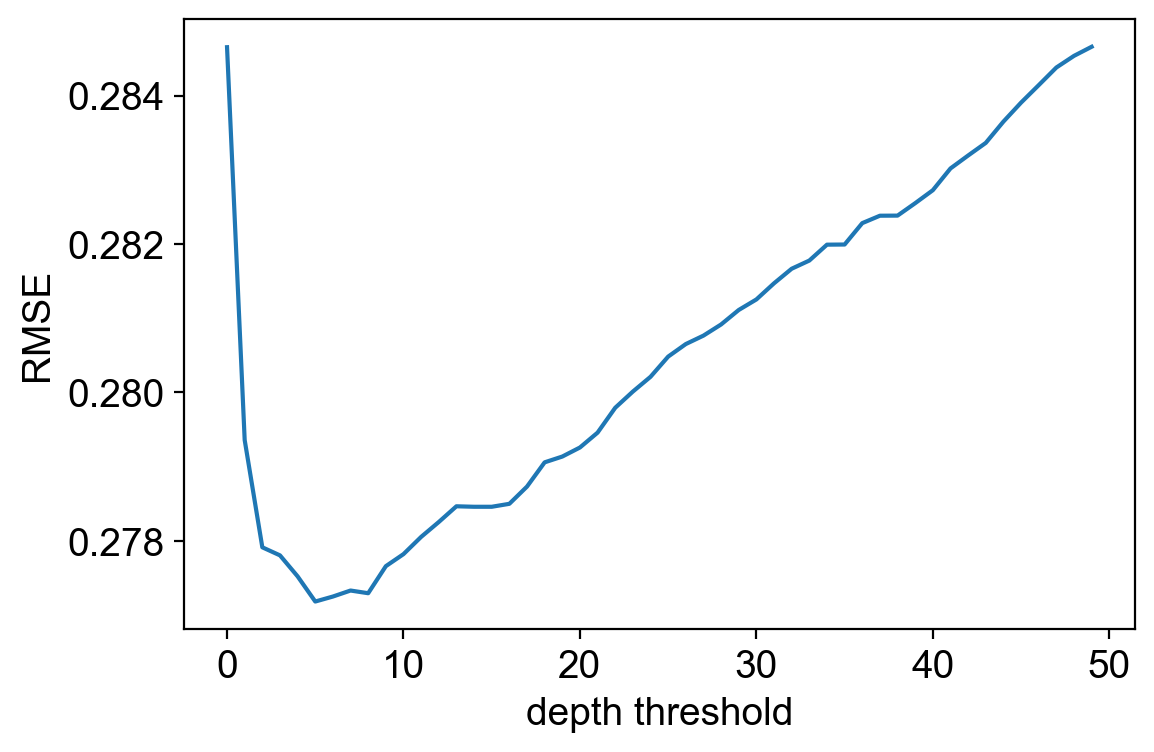}
    \includegraphics[width=0.335\linewidth]{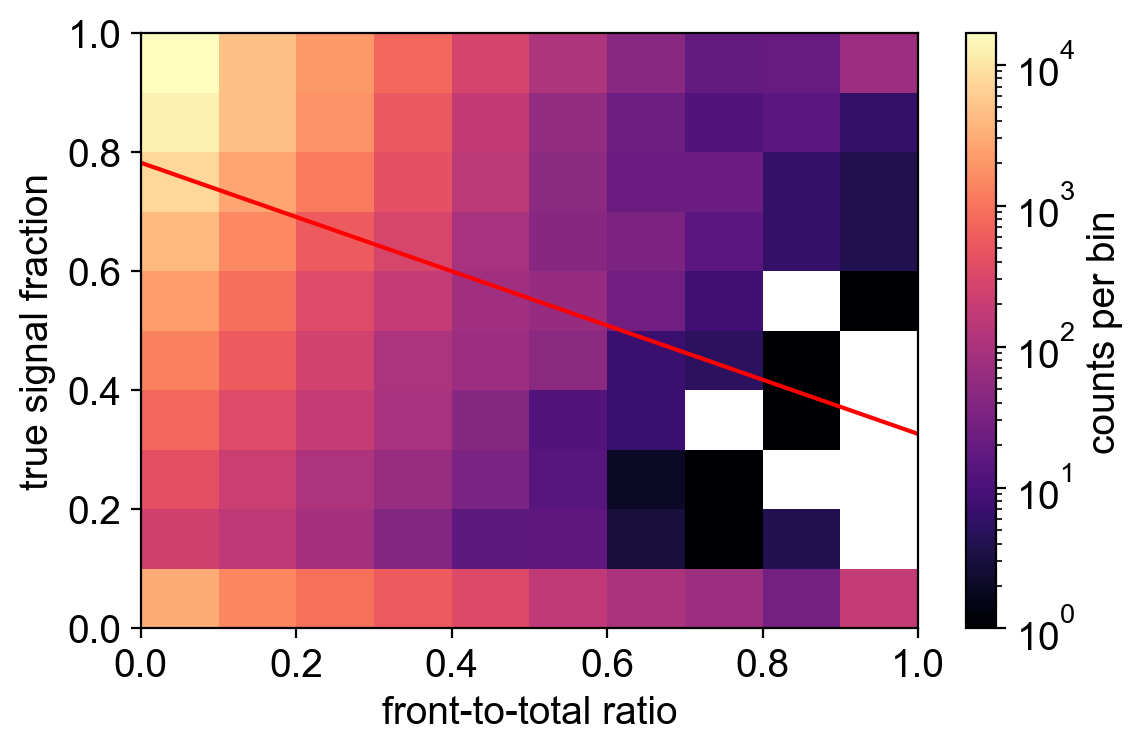}
    \includegraphics[width=0.290\linewidth]{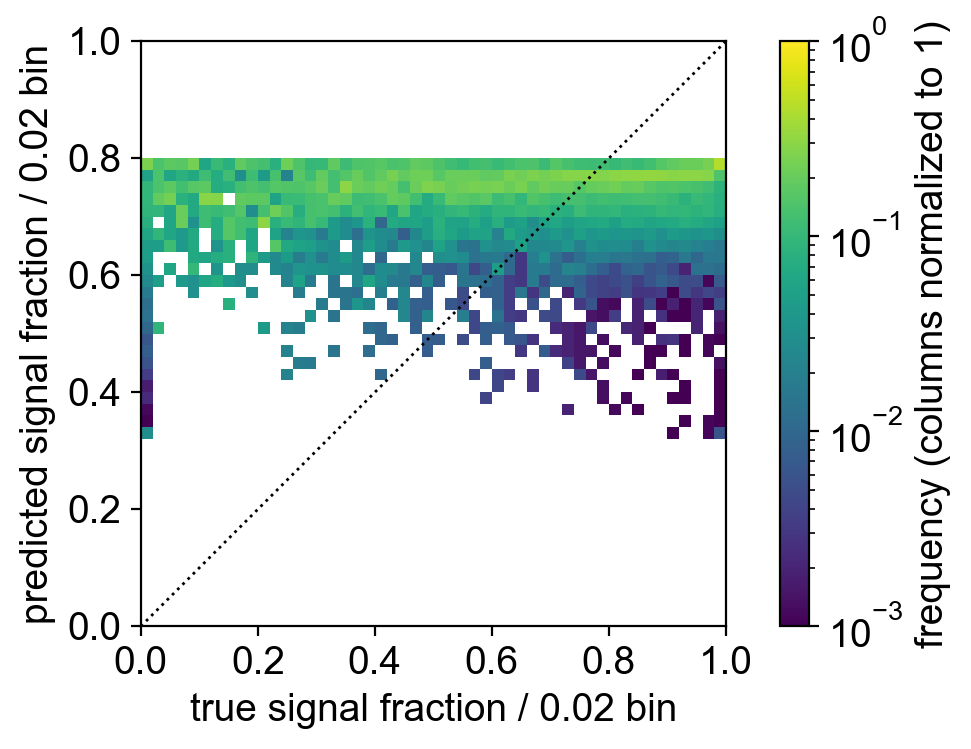}
    \caption{
        Results for the classical linear regression model.
        (a) RMSE between predicted $\hat{f}$ and true signal fraction $f$ of the validation set as a function of depth threshold $d_0$. 
        (b) Histogram of front-to-total ratio $R_{d_0}$ versus $f$ for the best performing threshold, bin 5, overlaid with the line of best fit from the linear regression plotted in red.
        (c) Response matrix of the linear model on the validation set.
    }
    \label{fig:regression}
\end{figure*}

Results for the linear regression model are shown in Fig.~\ref{fig:regression}.
The optimal threshold $d_0$, as determined by RMSE between the predicted signal fraction~$\hat{f}$ and true signal fraction $f$, was found to be bin 5.
Such a shallow bin results in the majority of signal-containing pulses having a $R_{d_0}$ of $0$.
The range of the fit was also quite narrow, predicting values only between roughly $0.35$--$0.80$.
In all, the linear regression model did not fit the data well, with the best RMSE attained being $27.7\%$.

\begin{figure*}[!htbp]
    \centering
    \includegraphics[width=0.335\linewidth]{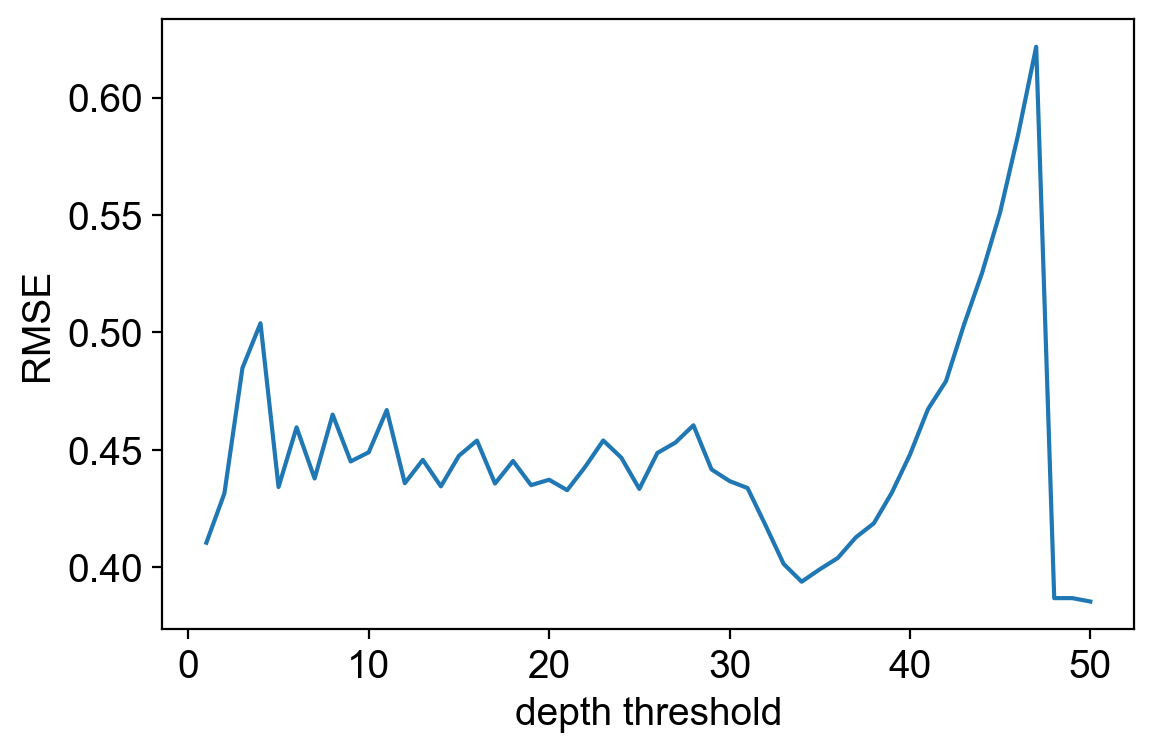}\label{fig:maxlike_a}
    \includegraphics[width=0.335\linewidth]{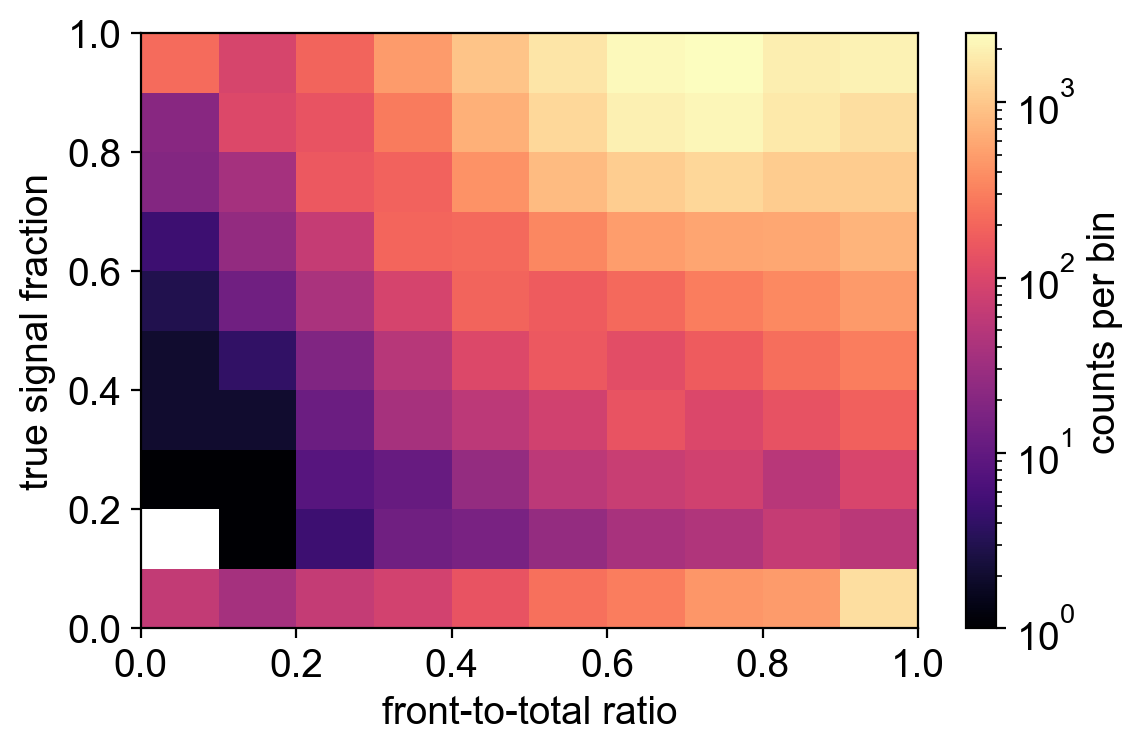}
    \includegraphics[width=0.290\linewidth]{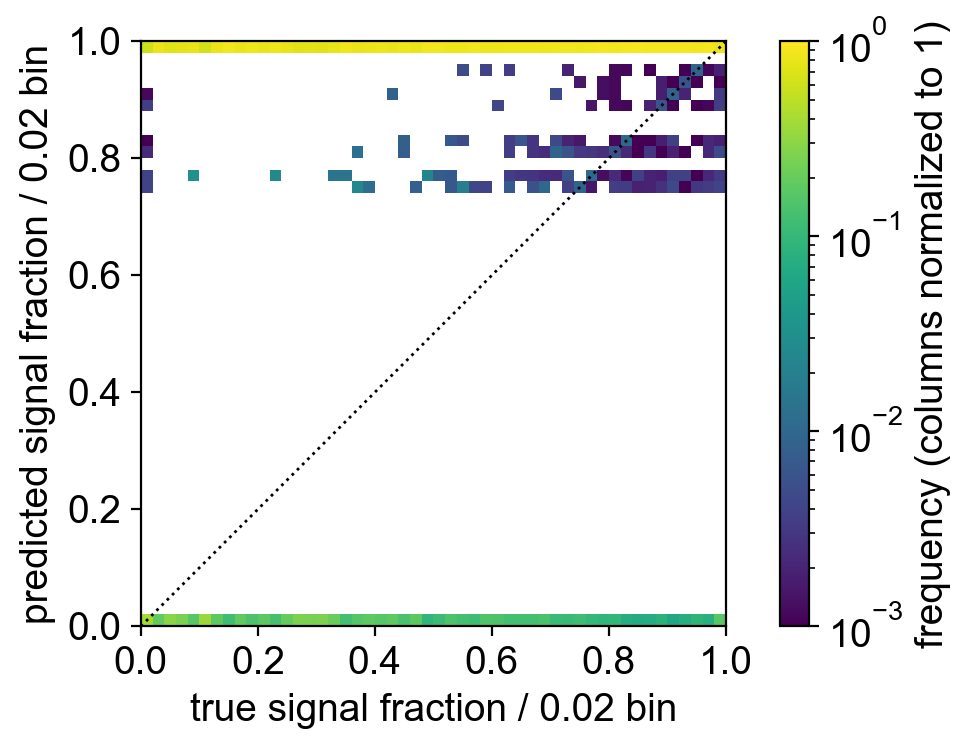}
    \caption{
        Results for the classical likelihood model with threshold $d_0$ set at bin 34. (a) RMSE between predicted $\hat{f}$ and true signal fraction $f$ of the validation set as a function of $d_0$. (b) Response matrix of the likelihood model on the validation set.
    }
    \label{fig:maxlike}
\end{figure*}

Fig.~\ref{fig:maxlike} shows the results for the maximum likelihood model.
The RMSE is minimized with a very deep threshold, bin 48, which, like the linear model above, results in the model predicting a constant $\hat{f}$.
Results in Fig.~\ref{fig:maxlike} are for the more instructive case with a threshold at bin 34, which achieves an RMSE of $40.7\%$, similar to the constant case but the data are better separated by the $R_{d_0}$.
While there are some some small number of pulses on the response matrix that have predictions clustered around the $1{:}1$ diagonal, indicating accurate predictions of $\hat{f}$, the vast majority of predictions are for pure signal or background regardless of true signal fraction.

\begin{figure*}[!htbp]
    \centering
    \includegraphics[width=\linewidth]{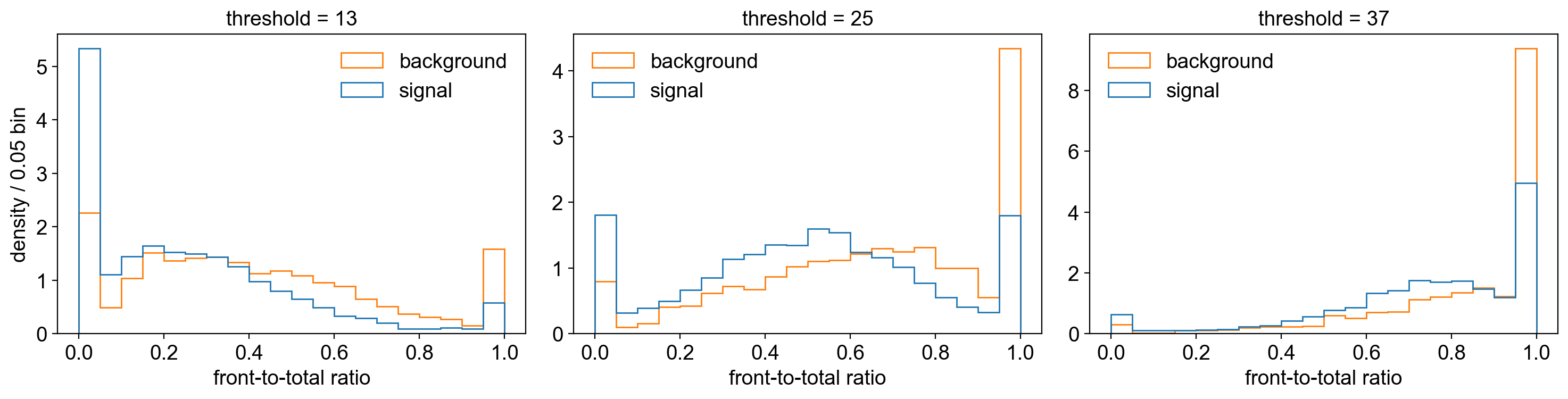}
    \caption{
        Front-to-total energy deposition ratios, $R_{d_0}$, for pure signal pulses (blue) and pure background pulses (orange) for three depth thresholds, (a) 13 bins (b) 25 bins, and (c) 37 bins in a 50 depth bin detector.
    }
    \label{fig:depth_threshold}
\end{figure*}

A closer examination of $R_{d_0}$ for pure signal and pure background interactions (Fig.~\ref{fig:depth_threshold}) shows that signal and background interactions can produce a wide range of $R_{d_0}$ values, and that the distributions of the two have substantial overlap.
As such, models relying on the simple depth ratio to separate events will be of limited utility.

\subsection{Multiplicity recovery with synthetically-piled-up real data}\label{sec:results_real_data}

\begin{figure}[!htbp]
    \centering
    \includegraphics[width=1.0\linewidth]{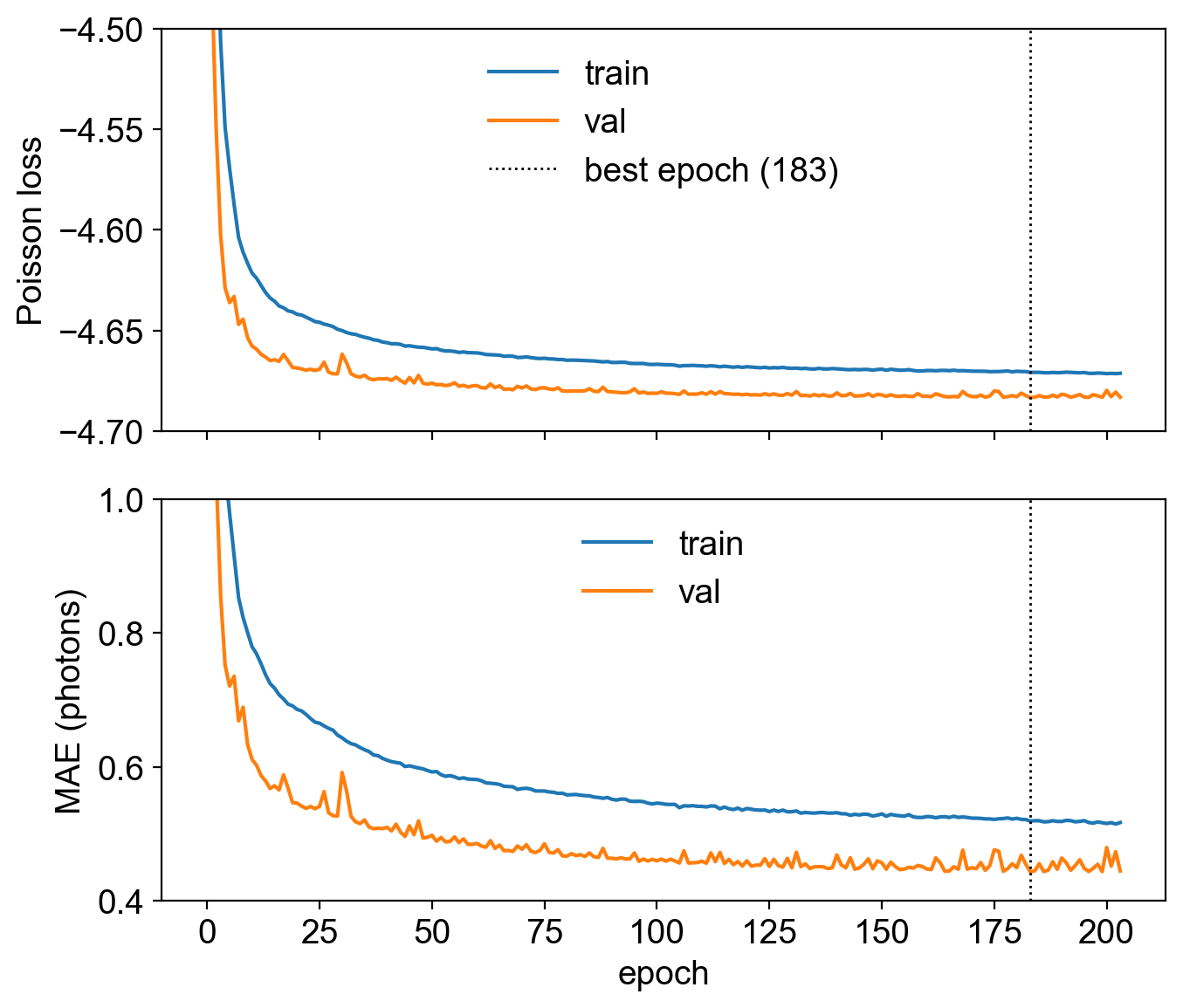}
    \includegraphics[width=1.0\linewidth]{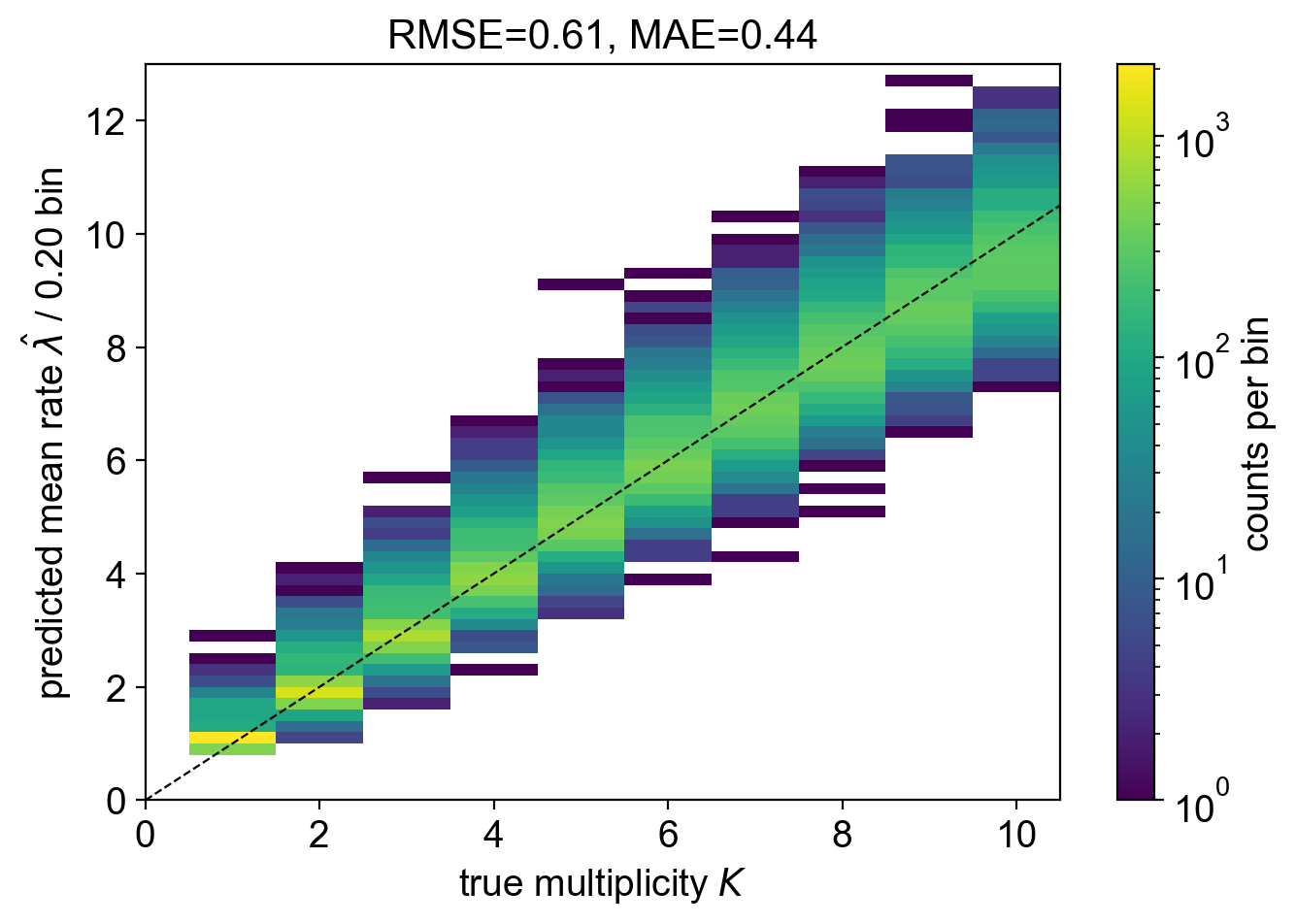}
    \caption{
        Real-data multiplicity recovery results.
        Top: Poisson loss vs.\ epoch.
        Middle: MAE vs.\ epoch.
        Bottom: response matrix between predicted Poisson mean rate~$\hat{\lambda}$ and true multiplicity~$K$.
    }
    \label{fig:real_data}
\end{figure}

Fig.~\ref{fig:real_data} shows results from the multiplicity recovery problem using the synthetically-piled-up Cs-137 M400i data.
The CNN achieves a minimum validation Poisson loss at epoch~$183$ (0-indexed), corresponding to a mean absolute error (MAE) in multiplicity of $0.44$ photons (RMSE of $0.61$).

\begin{figure}[!htbp]
    \centering
    \includegraphics[width=1.0\linewidth]{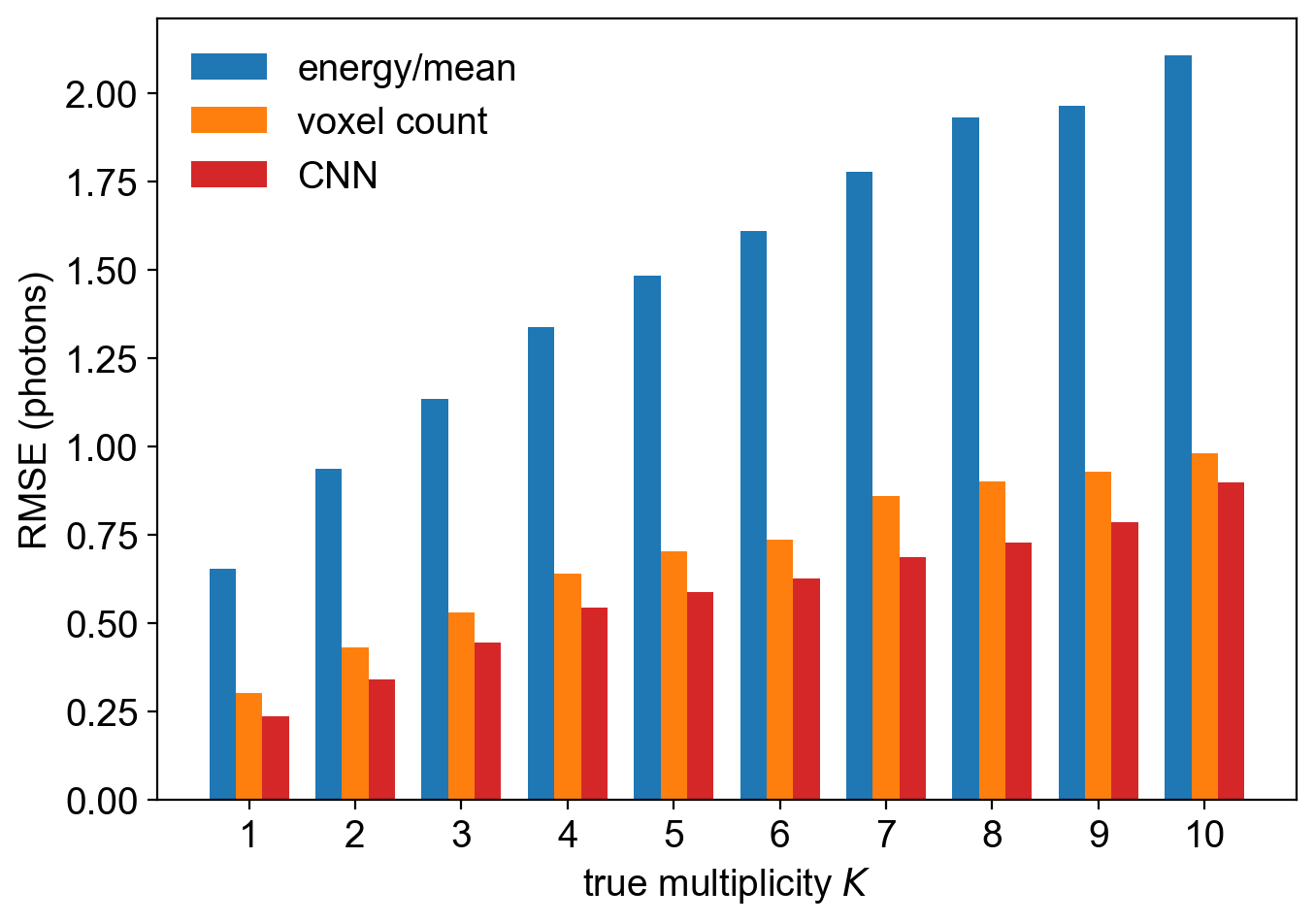}
    \caption{
        RMSE vs.\ true multiplicity $K$ for the CNN and the two baseline (non-ML) methods.
    }
    \label{fig:real_data_rmse_vs_k}
\end{figure}

Fig.~\ref{fig:real_data_rmse_vs_k} shows the trends in RMSE vs.\ the true multiplicity $K$ for the 3D CNN and the two baseline methods described in Section~\ref{sec:methods_meas_data}.
The CNN (overall RMSE of $0.61$) clearly outperforms the total energy method (which simply divides the total energy deposition by the mean expected deposition; RMSE of $1.56$), and modestly outperforms the voxel count method (which divides the number of hit voxels by the mean expected number of hit voxels for $K=1$ photon; RMSE of $0.74$).

\section{Discussion}\label{sec:discussion}
We have shown via two example problems that various neural network architectures can be used to recover spectroscopic information that would be lost to ultrafast pileup via spatial energy deposition patterns.
Here we provide some additional discussion, including limitations and opportunities for future work.

First, it is worth emphasizing that in the signal fraction recovery problem, the classical, non-ML methods in general do not perform well---as expected---achieving minimum RMSEs of $28\%$ (linear regression) and $41\%$ (maximum likelihood).
Performance here is fundamentally limited by the simplicity of these essentially heuristic methods, and the fact that they use only depth information, and will not improve with additional data as the ML-based methods will.

Second, we remark that the FCNN architecture, which does not inherently preserve the 3D structure of the data, can achieve an RMSE slightly worse than, but still similar to, that of the inherently-3D CNN ($18.8\%$ vs.\ $14.5\%$).
It is possible that the FCNN essentially relearns the 3D structure, and/or that the kernel sizes considered for the CNN exclude important 3D correlations that would further improve the CNN performance relative to the FCNN.
ML explanation methods such as Integrated Gradients~\cite{sundararajan2017axiomatic} could provide more insight here.
Related, hyperparameter optimization has only been minimally explored for the CNN architecture, and could be valuable in the future.
It is also interesting to note that the 3D CNN alone outperforms the FCNN equipped with hyperparameter optimization and the PointNet++ model equipped with weighted sampling, again suggesting that the 3D CNN is the appropriate architecture.

Third, real data may present additional challenges not considered in synthetic models; while our multiplicity recovery demonstration used measured data for both the training and validation stages, future applications may be limited to training on synthetic data and testing on real data.
Pixel- or voxel-level resolution and efficiency variations, as well as dead pixels, may negatively impact spectroscopic recovery performance if not accounted for in the training datasets.
To this end, it could be interesting to couple the present neural network methods with the data-driven spectroscopic optimization methods of Ref.~\cite{vavrek2026data}, using the neural network RMSE as an optimization target.

Fourth, while we have modeled perfect discretization in all three detector dimensions for our proofs of concept, real CZT detectors using pixelated anodes plus depth-of-interaction estimation cannot readily distinguish two simultaneous different-depth interactions within the same pixel, so our results are likely optimistic.
It would therefore be valuable in future work to incorporate this information loss mechanism into our synthetic models and quantify its impact on information recovery performance.
If performance degrades substantially, it could be interesting to explore rotating the detector by $90^\circ$ such that the anode segmentation lies in the beam direction, potentially sacrificing fidelity in one transverse coordinate for more reliable depth estimation, which may be suitable or even preferential for networks that operate primarily on depth information.

Finally, we emphasize again that the problems of recovering the signal fraction~$f$ and the multiplicity~$K$ are representative proofs of concept for ultrafast information recovery methods.
Future dedicated applications may require different quantities to be reconstructed and will likely use different, non-ideal detector configurations.

\section{Conclusion}
We have shown that machine learning methods can be used to recover spectroscopic information lost due to ultrafast pileup.
These ML methods exploit the spatial distribution of photon hit patterns in position-sensitive detectors, rather than temporal information.
3D convolutional neural networks tend to perform best in our example problems, but other network architectures can often achieve similar performance.
In the future, these methods could re-enable spectroscopic analyses in a broad array of photon active interrogation applications with fast pulsed sources and slow detectors.

\appendices

\renewcommand\thefigure{\thesection.\arabic{figure}}
\setcounter{figure}{0}
\renewcommand\thetable{\thesection.\Roman{table}}
\setcounter{table}{0}

\section{Number of possible event sequences for $N$ hits}\label{sec:appendix_sequences}
\begin{figure}[!htbp]
    \centering
    \includegraphics[width=1.0\columnwidth]{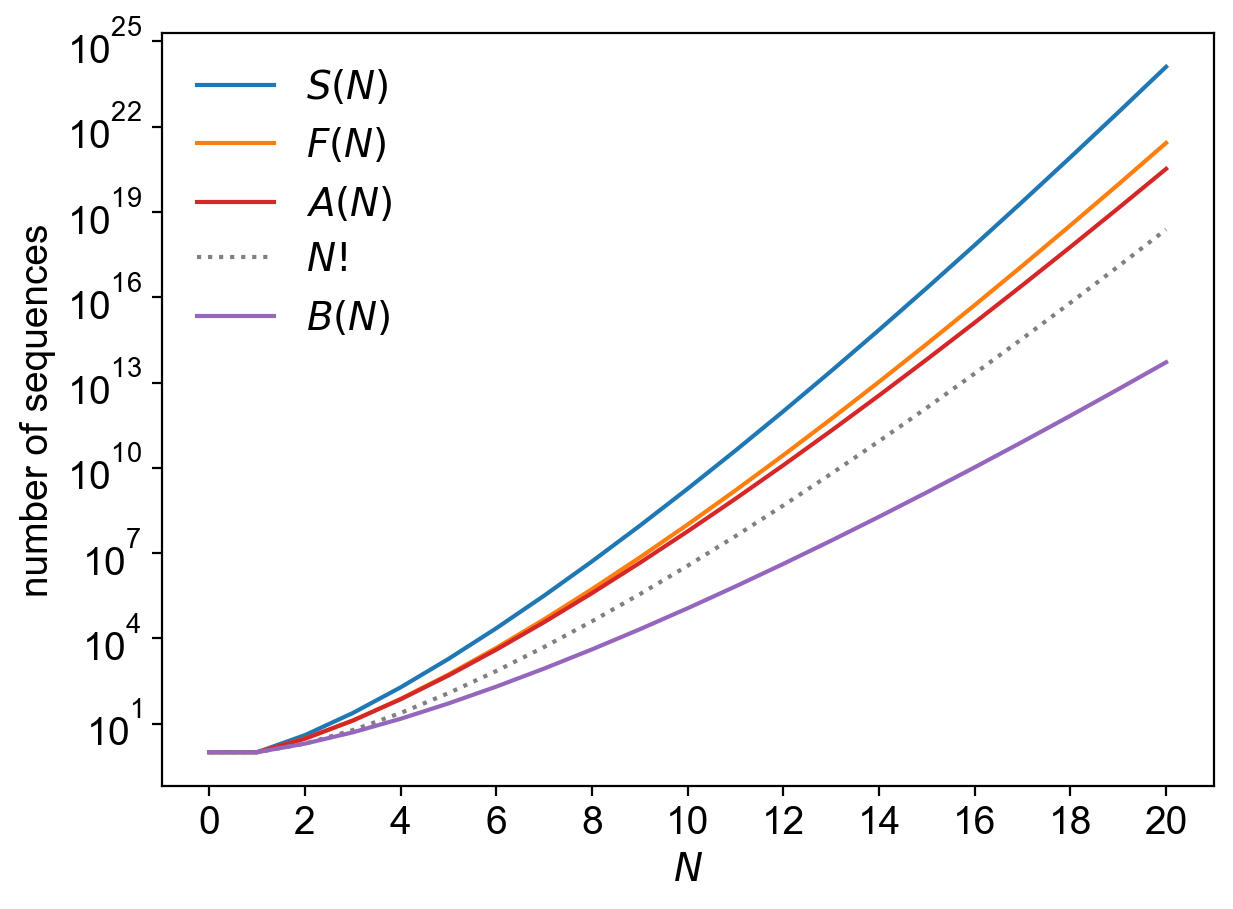}
    \caption{
        Number of interaction sequences given $N$~hits.
        Although~$N$ takes whole-number values, continuous curves are drawn for visual clarity.
    }
    \label{fig:n_event_sequences}
\end{figure}

In the ultrafast pileup problem, $N$ photon interactions or ``hits'' may stem from $1 \leq K \leq N$ incident photons piled up within the ultrafast pulse, where the true number of incident, interacting photons $K$ is not known.
Generally, assigning $N$ hits to $k = 1, \ldots, K$ groups can be accomplished in four different ways depending on to what degree hit order matters, equivalent to enumerating the number of collections of collections, where, adopting terminology from Python, a collection can be a list (order matters) or a set (order does not matter):
\begin{enumerate}
    \item the number of sequences where both intra- and inter-event sequence information matter, denoted~$S(N)$, is equivalent to enumerating the number of lists of lists of length~$N$;
    \item the number of sequences where inter-event ordering matters but intra-event ordering does not is given by the Fubini or ordered Bell numbers, denoted~$F(N)$, and is equivalent to enumerating the number of lists of sets;
    \item the number of sequences where intra-event ordering matters but inter-event ordering does not, denoted~$A(N)$, is equivalent to enumerating the number of sets of lists;
    \item the number of sequences where neither inter- nor intra-event ordering matters is given by the Bell numbers, denoted~$B(N)$, and is equivalent to enumerating the number of sets of sets.
\end{enumerate}
Under traditional maximum likelihood methods based on Klein-Nishina physics, a complete reconstruction would involve $S(N)$ possible likelihood evaluations.
A more typical analysis would fall into the third class, scaling as $A(N)$, evaluating every possible intra-event ordering but discarding the inter-event ordering information.
An analysis that only computes all possible energy deposition combinations without kinematic (ordering) information would fall into the fourth class, requiring $B(N)$ combinations.
$F(N)$ is included only for completeness.
Fig.~\ref{fig:n_event_sequences} shows the sequences $S(N)$, $F(N)$, $A(N)$, and $B(N)$ up to $N=20$.
The first three grow faster than $N!$, and $B(N)$ grows only somewhat more slowly than $N!$, indicating that brute-force testing all possible sequences quickly becomes expensive even at modest values ($N \gtrsim 10$).
For instance, if a likelihood could be evaluated in as low as $1$~{\textmu}s, a single collection of $N=10$ hits would require ${\sim}1$~core-minute at $A(N)$ scaling, while $N=15$ and $20$ would require ${\sim}2$ and ${\sim}10^7$~core-years, respectively.
Most measurements would involve many individual event collections, scaling the total compute time further.
Table~\ref{tab:n_event_sequences} provides additional summary info.

\begin{table*}[!htbp]
    \centering
    \caption{\textsc{Number of possible event sequences with various information loss mechanisms}}
    \begin{tabular}{c|c|c|c|c}
         & exact reconstruction & intra-event loss & inter-event loss & both losses \\\hline
         symbol & $S(N)$ & $F(N)$ & $A(N)$ & $B(N)$ \\
         $N=0,\ldots,5$ & $1, 1, 4, 24, 192, 1920$ & $1, 1, 3, 13, 75, 541$ & $1, 1, 3, 13, 73, 501$ & $1, 1, 2, 5, 15, 52$ \\
         name & n/a & Fubini / ordered Bell & n/a & Bell \\
         OEIS* & A002866~\cite{OEIS_A002866} & A000670~\cite{OEIS_A000670} & A000262~\cite{OEIS_A000262} & A000110~\cite{OEIS_A000110} \\
         equivalence & \# lists of lists  & \# lists of sets & \# sets of lists & \# sets of sets \\
    \end{tabular}
    \label{tab:n_event_sequences}
    \\\vspace{2pt}
    *On-line Encyclopedia of Integer Sequences
\end{table*}

\section{Uncertainty quantification for the CNN}\label{sec:appendix_uq}

In the main text, we had only considered point estimates for our reconstructed spectroscopic quantities, without any uncertainty quantification (UQ).
Here, we include a small demonstration of UQ for the CNN signal fraction recovery problem using the Monte Carlo (MC)-Dropout method~\cite{gal2016dropout}, which applies to any network already trained with dropout.
In brief, randomly sampling the dropout mask at inference time provides approximate posterior information that can be used to compute (often uncalibrated) estimates of the posterior epistemic (reducible) uncertainty.
Aleatoric (irreducible) uncertainty is not addressed by this method.

\begin{figure*}[!htbp]
    \includegraphics[width=1.0\textwidth]{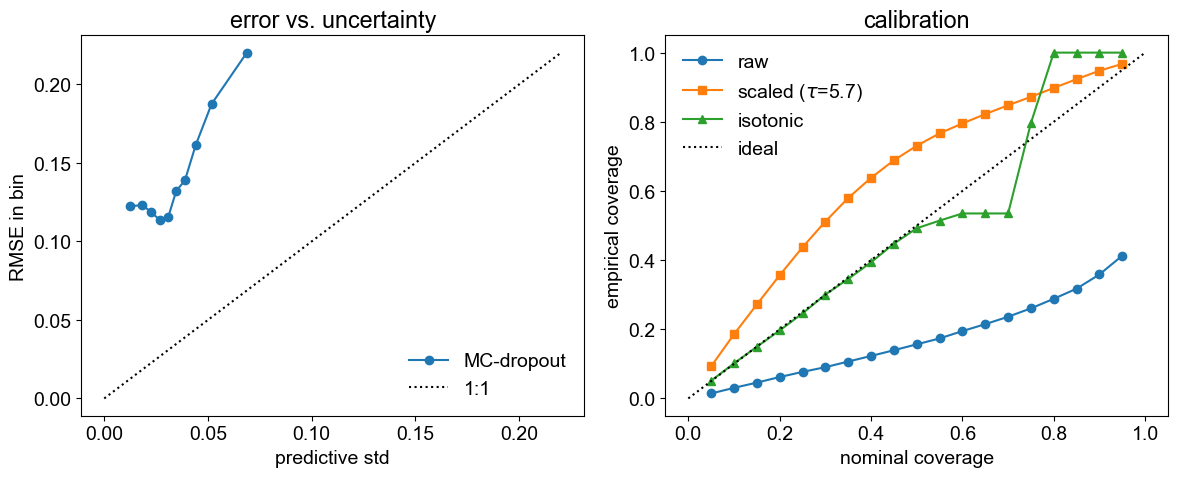}
    \caption{
        MC-dropout UQ results for the signal fraction recovery CNN.
        Left: RMSE between the true and predicted signal fractions $f$ and $\hat{f}$, binned by the MC-dropout standard deviation in $\hat{f}$ values.
        Right: empirical vs.\ nominal coverage for the raw, globally-scaled, and isotonic-scaled uncertainty estimates.
    }
    \label{fig:uq_mcdropout}
\end{figure*}

Fig.~\ref{fig:uq_mcdropout} shows the MC-dropout results using $T=200$ random samples per (validation-set) pulse.
For each pulse, the standard deviation in predicted $\hat{f}$ (the ``predictive standard deviation'') is used as a raw, uncalibrated uncertainty estimate.
Pulses are then binned by predictive standard deviation, and the RMSE of the predicted $\hat{f}$ values (the ``observed error'') around their respective true $f$ is computed in each bin.
An ideal UQ method would lie on the $1{:}1$ line between the predictive standard deviation and the observed error.
Our RMSE values are consistently higher than the $1{:}1$ line, indicating the predictive standard deviation estimates underpredict the observed error in $\hat{f}$ about the true value $f$.
While the raw uncertainties are overconfident in scale (mean or ``sharpness'' of $0.035$), they are generally informative, i.e., they correlate positively with RMSE.
Similarly, the empirical vs.\ nominal coverage curve for the raw uncertainty estimates is consistently below the ideal $1{:}1$ line, indicating that the predictive standard deviation in $\hat{f}$ produces uncertainty intervals that contain the true $f$ less often than nominally expected.

This miscalibration of raw uncertainties is common for deep networks and several methods for recalibration have been proposed; in particular we compare both a global rescaling~\cite{levi2022evaluating} and isotonic recalibration~\cite{kuleshov2018accurate} of the raw uncertainties.
In both cases, the recalibrations are computed on one half of the validation set and evaluated on the other.
Calibration quality before and after recalibration is measured through the average calibration error~\cite{psaros2023uncertainty}, i.e., the mean absolute deviation of each coverage curve with respect to the diagonal in Fig.~\ref{fig:uq_mcdropout}.
The global method rescales the predictive standard deviations by a single scalar $\tau$ that matches the average predictive standard deviation to the observed error standard deviation.
We find that $\tau = 5.7$ reduces the average calibration error from the raw $0.33$ to $0.15$, and gives a rescaled mean predictive uncertainty of ${\sim}0.2$, but still retains significant non-linearity in the empirical vs.\ nominal coverage calibration.
The isotonic rescaling performs better with an average calibration error of $0.05$, and is nearly $1{:}1$ until a nominal coverage of about $0.5$.

Roughly speaking, the raw uncertainties are overconfident by the average factor of $\tau = 5.7$ computed via the global rescaling, and there is variation in this overconfidence vs.\ nominal coverage that is better captured by the isotonic correction.
After the global rescaling, the mean uncertainty of the CNN signal fraction output is ${\sim}0.20$, which is wide given the $[0, 1]$ range of signal fractions~$f$.
Applying dropout after ``every network unit in each layer'' as per the original formulation in Ref.~\cite{gal2016dropout} may allow for richer posterior sampling to help narrow the calibrated uncertainties, and would be a valuable direction to test in future work.

The same MC-dropout method could be extended to future versions of our existing FCNN models that were augmented with dropout layers.
The PointNet++ model would likely require a more architecture-agnostic UQ method such as deep ensembles~\cite{lakshminarayanan2017simple}, though this would require multiple expensive training runs, not just multiple prediction runs as with MC-dropout.

\section*{Acknowledgment}
\small
The authors thank Hannah Parrilla (LBNL) for acquiring the Cs-137 data, and Jeroen van Tilborg, Cameron Geddes, Mark Bandstra, and Brian Quiter (LBNL) for useful discussions.

Early versions of the CNN were prototyped by an OpenAI GPT model available in November 2024, most likely a GPT 4-series.
Claude Opus 4.6 was used for the measured data postprocessing and visualization in Section~\ref{sec:results_real_data}.
The UQ demonstration in Appendix~\ref{sec:appendix_uq} was designed and implemented via Claude Opus 4.8 (1M context).
The manuscript text was written by humans and reviewed by Claude Opus 5 (1M context).

This manuscript has been authored by an author at Lawrence Berkeley National Laboratory under Contract No.\ DE-AC02-05CH11231 with the U.S.\ Department of Energy.
The U.S.\ Government retains, and the publisher, by accepting the article for publication, acknowledges, that the U.S.\ Government retains a non-exclusive, paid-up, irrevocable, world-wide license to publish or reproduce the published form of this manuscript, or allow others to do so, for U.S.\ Government purposes.
The views and opinions of authors expressed herein do not necessarily state or reflect those of the United States Government or any agency thereof or the Regents of the University of California.

\bibliographystyle{IEEEtran}
\bibliography{biblio}

\end{document}